\documentclass[preprint,12pt]{elsarticle}

\usepackage{lineno,hyperref}
\modulolinenumbers[5]

\usepackage{graphicx}
\usepackage{comment}
\usepackage{xspace}
\usepackage{amssymb}

\newcommand{\Math}[1]{\ensuremath{#1}}

\newcommand{\modecal}[1]{{\Math{\mathcal{#1}}}}

\newcommand{\textmath}[1]{\mbox{\textit{#1}}}

\newcommand{\propername}[1]{\mbox{\small \textsf{#1}}\xspace}

\newcommand{\Golog}{\propername{Golog}}

\newcommand{\ConGolog}{\propername{ConGolog}}

\newcommand{\A}{\modecal{A}} 
\newcommand{\C}{\modecal{C}} \newcommand{\D}{\modecal{D}}
 \newcommand{\F}{\modecal{F}}

\newcommand{\M}{\modecal{M}} \newcommand{\N}{\modecal{N}}
 
\renewcommand{\S}{\modecal{S}}

\newcommand{\R}{\modecal{R}} 

\newcommand{\ndet}{\mbox{$|$}}
\newcommand{\conc}{\mbox{$\parallel$}}

\newcommand{\Trans}{\textmath{Trans}}

\newcommand{\Poss}{\textmath{Poss}}

\def\planeaux!#1:#2<-#3!{\Math{#1 \mbox{\rm:} #2\; \leftarrow #3}}

\def\planeaux!#1<-#2!{\Math{#1 \leftarrow #2}}

\newcommand{\Nat}{\Math{\mathbb{N}}}

\long\def\eatpar#1{%
\ifx#1\par                      
\let\nextmove=\eatpar           
\else
\let\nextmove=#1
\fi
\nextmove
}

\def\qed{\hfill{\qedboxempty}      
  \ifdim\lastskip<\medskipamount \removelastskip\penalty55\medskip\fi}

\def\qedboxempty{\vbox{\hrule\hbox{\vrule\kern3pt
                 \vbox{\kern3pt\kern3pt}\kern3pt\vrule}\hrule}}

\def\qedfull{\hfill{\qedboxfull}   
  \ifdim\lastskip<\medskipamount \removelastskip\penalty55\medskip\fi}

\def\qedboxfull{\vrule height 4pt width 4pt depth 0pt}

\newcounter{bean}

\newenvironment{tightenumerate}{
                \begin{list}{
                  {\mbox {
                      \arabic{bean}.\/}}}{\usecounter{bean}
                      \setlength{\itemsep}{-1pt}\setlength{\topsep}{0pt}}}{
                \end{list}}

\newenvironment{tightitemize}{
                \begin{list}{$\bullet$}{
                    \setlength{\itemsep}{-1pt}}{\setlength{\topsep}{0pt}}}{
                \end{list}}

\newcommand{\under}[1]{\mbox{\underline{\it\smash{#1}\vphantom{\lower.05ex\hbox{
x}}}}}

\newcommand{\commentarea}[1]{}

\newtheorem{thm}{Theorem}
\newtheorem{lem}[thm]{Lemma}
\newdefinition{rmk}{Remark}
\newproof{pf}{Proof}
\newproof{pot}{Proof of Theorem \ref{thm2}}

\newtheorem{corollary}[thm]{Corollary}
\newtheorem{definition}[thm]{Definition}  
\newtheorem{proposition}[thm]{Proposition}

\newtheorem{exampleAux}{Example}

\newtheorem{constraint}{Constraint}

\newcommand{\SitDet}{\mathit{SituationDetermined}}
\newcommand{\next}{\mathit{next}}
\newcommand{\limp}{\supset}

\newcommand{\CGaxioms}{\mathcal{C}}

\newcommand{\anyonehl}{\textsc{any1hl}}
\newcommand{\anyseqhl}{\textsc{anyseqhl}}

\usepackage{xcolor}

\newcommand*{\qedf}{\hfill\ensuremath{\blacksquare}}

\journal{Artificial Intelligence}

\begin{document}

\begin{frontmatter}

\title{Abstracting Situation Calculus Action Theories}

\author[1]{Bita Banihashemi\corref{cor1}}
\ead{bita.banihashemi@igdore.org}
\author[2,3]{Giuseppe De Giacomo}
\ead{giuseppe.degiacomo@cs.ox.ac.uk}
\author[4]{Yves Lesp\'{e}rance}
\ead{lesperan@eecs.yorku.ca}

\cortext[cor1]{Corresponding author}
\address[1]{IGDORE, Gothenburg, Sweden}
\address[2]{University of Oxford, Oxford, United Kingdom}
\address[3]{Sapienza University of Rome}
\address[4]{York University, Toronto, Canada}






\begin{abstract}
We develop a general framework for \emph{agent abstraction} based 
 on the situation calculus and the \ConGolog agent programming language.
We assume that we have a high-level specification and a low-level
specification of the agent, both represented as basic action
theories. A \emph{refinement mapping} specifies how each high-level
action is implemented by a low-level \ConGolog program and how each
high-level fluent can be translated into a low-level formula.  We
define a notion of \emph{sound abstraction} between such action
theories in terms of the existence of a suitable bisimulation between
their respective models.  Sound abstractions have many useful
properties that ensure that we can reason about the agent's actions
(e.g., executability, projection, and planning) at the abstract level,
and refine and concretely execute them at the low level.  
We also characterize the notion of \emph{complete
  abstraction} where all actions (including exogenous ones)  that the high level thinks can happen
can in fact occur at the low level. 
To facilitate verifying that one has a sound/complete abstraction relative to a mapping, we provide a set of necessary and sufficient conditions.
Finally, we identify a set of basic action theory constraints that ensure that for any low-level action
sequence, there is a unique high-level action sequence that it refines. This allows us to track/monitor what the low-level agent is doing and describe it in abstract terms (i.e., provide high-level explanations, for instance, to a client or manager).

\end{abstract}



\end{frontmatter}



\section{Introduction}

Autonomous agents often operate in complex domains and have complex
behaviors.\footnote{This work revises and extends \cite{DBLP:conf/aaai/BanihashemiGL17}.}
Reasoning about such agents and even describing their behavior can be difficult.  
One way to cope with this is to use \emph{abstraction} \cite{Saitta:Zucker:2013}.  In essence, this involves developing an abstract model of the agent/domain that suppresses less important details.
The abstract model allows us to reason more easily about the agent's possible behaviors and to provide high-level explanations of the agent's behavior.   
To efficiently solve a complex reasoning problem, e.g.~planning, one may first try to find a solution in the abstract model, and then use this abstract solution as a template to guide the search for a solution in the concrete model.
%
%
Systems developed using abstractions are typically more robust to change, as adjustments to more detailed levels may leave the abstract levels unchanged.

In this paper,
we develop a general framework for \emph{agent abstraction} based on the 
situation calculus \cite{McCarthy1969:AI,Reiter01-Book} and the \ConGolog \cite{DBLP:journals/ai/GiacomoLL00} agent programming language.
We assume that one has a high-level/abstract action theory, a low-level/concrete action theory, and a \emph{refinement mapping} between the two.  The mapping associates each high-level primitive action to a (possibly nondeterministic) \ConGolog program defined over the low-level action theory that ``implements it''.  Moreover, it maps each high-level fluent to a state formula in the low-level language that characterizes the concrete conditions under which it holds.

In this setting, we define a notion of a high-level theory being a \emph{sound abstraction} of a low-level theory under a given refinement mapping.  
The formalization involves the existence of a suitable bisimulation relation \cite{DBLP:conf/ijcai/Milner71,DBLP:books/daglib/0067019} 
relative to a mapping 
between models of the low-level and high-level theories.
With a sound abstraction, whenever the high-level theory \emph{entails} that a sequence of actions is executable and achieves a  certain condition, then the low level must also entail that there exists an executable refinement of the sequence such that the ``translated'' condition holds afterwards. Moreover, whenever the low level thinks that a refinement of a high-level action 
(perhaps involving exogenous actions) 
can occur (i.e., its executability is satisfiable), then the high level does as well. 
Thus, sound abstractions can be used to perform effectively several forms of reasoning about action, such as planning, agent monitoring, and generating high-level explanations of low-level behavior.  
%

In addition, we define a dual notion of \emph{complete abstraction}
where whenever the low-level theory \emph{entails} that some
refinement of a sequence of high-level actions is executable and
achieves a ``translated'' high-level condition, then the high level
also \emph{entails} that the action sequence is executable and the
condition holds afterwards.  Moreover, whenever the high level thinks
that an action can occur (i.e., its executability is satisfiable),
then there exists a refinement of the action that the low level thinks
can happen as well.

We also provide a set of necessary and sufficient conditions for
having a sound and/or complete abstraction relative to a mapping.
These can be used to verify that that one has a sound/complete
abstraction.

Finally, we identify a set of basic action theory constraints that ensure that for any low-level action
sequence, there is a unique high-level action sequence that it refines. This allows us to track/monitor what the low-level agent is doing and describe it in abstract terms (i.e., provide high-level explanations \cite{DBLP:conf/aaai/Doyle86}) e.g., to a client or manager. This can have applications in \emph{Explainable AI}.

In the past, many different approaches to abstraction have been proposed in a
variety of settings such as planning \cite{DBLP:journals/ai/Sacerdoti74,DBLP:journals/amai/ErolHN96,Korf:1987}, automated
reasoning \cite{Giunchiglia:Walsh:1992,DBLP:conf/ijcai/NayakL95},
model checking \cite{DBLP:journals/toplas/ClarkeGL94}, and data integration \cite{DBLP:conf/pods/Lenzerini02}.
With the exception of work on hierarchical planning, these approaches
do not deal with dynamic domains.  Previous work on hierarchical
planning focuses on the planning task and often incorporates
important representational restrictions
\cite{DBLP:books/daglib/0014222,DBLP:books/cu/GNT2016}.
In contrast, our approach provides a generic framework that can be
applied to different reasoning tasks and deals with agents represented in an
expressive first-order framework.
We discuss related work in more details in Section \ref{sec:RelWork}.

The paper is organized as follows.
In the next section, we review the basics of the situation calculus and \ConGolog.
Then in Section \ref{sec:RefMap}, we define a notion of refinement
mapping between a high-level and a low-level basic action theory.
Section \ref{sec:Bisim} introduces our notion of bisimulation with respect to  a mapping that
relates models at the abstract and concrete levels.
Then in Sections \ref{sec:Sound} and \ref{sec:Complete}, we define the 
notions of sound and complete abstractions respectively, showing how
they allow the abstract theory to be exploited in reasoning.
In Section \ref{sec:MonExp}, we discuss how our framework can be used
in monitoring what the low-level agent is doing and explaining it in
abstract terms.
This is followed by a discussion of related work in Section
\ref{sec:RelWork}. 
In Section \ref{sec:extWork} we provide an overview of some of the adaptations and extensions to our abstraction framework. 
Finally in Section \ref{sec:Conclusion}, we conclude by summarizing
our main contributions and discussing future work.


\section{Preliminaries}\label{sec:preliminaries}

\paragraph{The Situation Calculus and Basic Action Theories}
%
%
%
%
The \textit{situation calculus} is a well known predicate logic language
for representing and reasoning about dynamically changing
worlds \cite{McCarthy1969:AI,Reiter01-Book}. 
All changes to the world are the result of \textit{actions},
which are terms in the language. 
%
A possible world history is represented by a term called a
\emph{situation}. The constant $S_0$ is used to denote the initial
situation.
Sequences of
actions are built using the function symbol $do$, such that $do(a,s)$
denotes the successor situation resulting from performing action $a$
in situation $s$.
%
Predicates and functions whose value varies from situation to situation are
called \textit{fluents}, and are denoted by symbols taking a
situation term as their last argument.
For example, we may have that $Door1$ is not open in the initial
situation $S_0$, i.e., $\lnot IsOpen(Door1,S_0)$ holds, but is open in the
situation that results from doing the action $open(Door1)$ in $S_0$,
i.e., $IsOpen(Door1,do(open(Door1),S_0)$ holds.
$s \sqsubset s'$ means that $s$ is a  predecessor situation of $s'$, and
$s \sqsubseteq s'$ stands for $s = s' \lor s\sqsubset s'$.
The abbreviation  $do([a_1, \ldots, a_n],s)$ stands for
$do(a_n,do(a_{n-1},\ldots , do(a_1,s) \ldots ))$; also
for an action sequence $\vec{a}$, we often write $do(\vec{a}, s)$ for $do([\vec{a}],s)$.
%


Within this language, one can formulate action theories that describe
how the world changes as a result of the available actions.  Here,
we concentrate on \emph{basic action theories} 
as proposed in \cite{PirriR:JACM99,Reiter01-Book}.
We also assume that there is a \emph{finite number of action types} $\A$.
Moreover, we assume that the terms of object sort are in fact a
countably infinite set $\N$ of standard names for which we have the unique
name assumption and domain closure.\footnote{This makes it easier to
  relate the high-level and low-level action theories.
One of the main consequences of assuming standard names is that quantification
can be considered substitutionally; for instance, $\exists x.P(x)$ is
true just in case $P(n)$ is true for some standard name $n$.}
For simplicity, and w.l.o.g., we assume that there are no functions
other than constants and no non-fluent predicates.
As a result, a basic action theory $\D$ is the union of the
following disjoint sets of first-order (FO) and second-order (SO) axioms: 
\begin{itemize}
	\item $\D_{S_0}$: (FO) \emph{initial situation description axioms} 
		describing the initial configuration of the world (such a description may be complete or incomplete);
              \item $\D_{poss}$: (FO) \emph{precondition axioms} of
                the form \[Poss(A(\vec x), s)\equiv \phi^{Poss}_A(\vec x,s),\]
                one per action type, stating the conditions
                $\phi^{Poss}_A(\vec x,s)$ under which an action $A(\vec x)$
                can be legally performed in situation $s$; these use a
                special predicate $Poss(a,s)$ meaning that action $a$
                is executable in situation $s$; $\phi_A(\vec x,s)$ is
                a formula of the situation calculus that is
                uniform in $s$;\footnote{A formula of of the situation
                  calculus 
                  is \emph{uniform} in $s$ if and only if it does not mention the predicates $Poss$ or $\sqsubset$, it does not quantify over variables of sort situation, it does not mention equality on situations, and whenever it mentions a term of sort situation in the situation argument position of a fluent, then that term is $s$ \cite{Reiter01-Book}.}

              \item $\D_{ssa}$: (FO) \emph{successor state axioms} of
                the form
                \[F(\vec x,do(a,s))\equiv\phi^{ssa}_F(\vec x,a,s),\]
                one per fluent, describing how the fluent changes when
                an action is performed; the right-hand side (RHS)
                $\phi^{ssa}_F(\vec x,a,s)$ is again a situation calculus
                formula uniform in $s$; successor state axioms encode
                the causal laws of the world being modeled; they take
                the place of the so-called effect axioms and provide a
                solution to the frame problem;

	\item $\D_{ca}$: (FO) unique name axioms for actions and (FO) domain closure on action types; 

	\item $\D_{coa}$: (SO) unique name axioms and domain closure for object constants in $\N$; 

	\item  $\Sigma$: (SO) foundational, domain independent, axioms
of the situation calculus~\cite{PirriR:JACM99}.
\end{itemize}
The abbreviation $Executable(s)$ is used to denote that every action performed in reaching situation $s$ was possible in the situation in which it occurred. When executability of situations is taken into consideration, we use $<$ instead of $\sqsubset$ to indicate precedence on situations; consequently, $s \leq s'$ indicates that $s'$ is a successor situation of $s$ and that every action between $s$ and $s'$ is in fact executable.

A key feature of BATs is the
existence of a sound and complete \textit{regression mechanism} for
answering queries about situations resulting from performing a sequence of actions
\cite{PirriR:JACM99,Reiter01-Book}.
In a nutshell, the regression operator $\R^*$ reduces a formula
$\phi$ about a particular future situation to an equivalent formula
$\R^*[\phi]$ about the initial situation $S_0$, essentially by
substituting fluent relations with the right-hand side formula of
their successor state axioms.
%
Another key result about BATs is the relative
satisfiability theorem \cite{PirriR:JACM99,Reiter01-Book}:
$\D$ is \emph{satisfiable} if and only if $\D_{S_0} \cup \D_{una}$
is satisfiable, the latter being a purely first-order theory (here,
$\D_{una}$ is the set of unique names axioms for actions).
This implies that we can check if a regressable formula $\phi$ is entailed by $\D$, by checking if its regression $\R^*[\phi]$ is entailed by  $\D_{S_0} \cup \D_{una}$ only.

\paragraph{High-Level Programs}

To represent and reason about complex actions or processes obtained by suitably
executing atomic actions, various so-called \emph{high-level programming
languages} have been defined.
Here we concentrate on (a fragment of) \ConGolog\ \cite{DBLP:journals/ai/GiacomoLL00} that includes the following constructs:

\begin{small}
\[ \delta ::=  \alpha  \mid  \varphi?  \mid  \delta_1;\delta_2  \mid   \delta_1 \ndet \delta_2  \mid  \pi x.\delta  \mid  \delta^*  \mid   \delta_1 \conc \delta_2   \]
\end{small}%
%
In the above, 
$\alpha$ is an action term, possibly with parameters, and $\varphi$ is a
situation-suppressed formula, i.e., a formula 
with all situation arguments in fluents suppressed. 
We denote by $\varphi[s]$ the
formula obtained from $\varphi$ by restoring the situation
argument $s$ into all fluents in $\varphi$. The test action $\varphi?$ checks if condition $\varphi$ holds in the current situation.
%
The sequence of program $\delta_1$ followed by program $\delta_2$ is denoted by $\delta_1; \delta_2$.
Program $\delta_1 \ndet \delta_2$ allows for the nondeterministic choice
between programs $\delta_1$ and $\delta_2$, while $\pi x.\delta$ executes program
$\delta$ for \textit{some} nondeterministic choice of a legal binding for
variable $x$ (observe that such a choice is, in general, unbounded).  $\delta^*$
performs $\delta$ zero or more times.
Program $\delta_1 \conc \delta_2$ expresses the concurrent execution
(interpreted as interleaving) of programs $\delta_1$ and $\delta_2$. We also use $nil$, an abbreviation
for $True?$, to represent the \emph{empty program}, i.e., when nothing remains to be performed.

Formally, the semantics of \ConGolog\ is specified in terms of
single-step transitions, using the following two predicates
\cite{DBLP:journals/ai/GiacomoLL00}: 
\emph{(i)} $Trans(\delta,s,\delta',s')$, which holds if one step
of program $\delta$ in situation $s$ may lead to situation $s'$ with
$\delta'$ remaining to be executed; and \emph{(ii)}  $Final(\delta,s)$, which
holds if program $\delta$ may legally terminate in situation $s$.
The definitions of $Trans$ and $Final$ we use are as in
\cite{DBLP:conf/kr/GiacomoLP10}, 
where 
the test construct $\varphi?$ does not yield
any transition, but is final when satisfied (see the appendix for details). 
Predicate $Do(\delta,s,s')$ means that program $\delta$, when executed starting in situation $s$, has as 
a legal terminating situation $s'$, and is defined as $Do(\delta,s,s') \doteq \exists \delta'. Trans^*(\delta,s,\delta',s') \land Final(\delta',s')$ where $\Trans^*$ denotes the reflexive transitive closure of $\Trans$.
In the rest, we use  $\CGaxioms$ to denote the axioms defining the \ConGolog\
programming language.

We say that \ConGolog program $\delta$ is \emph{situation-determined} (SD) in a situation $s$ \cite{DBLP:conf/aamas/GiacomoLM12} 
if and only if for every sequence of transitions, the remaining program is
determined by the resulting situation, i.e.,

\begin{small}
\[\begin{array}{l}
\SitDet(\delta,s) \doteq \forall s',\delta', \delta''. \\
\hspace{1em}\Trans^*(\delta,s, \delta',s') \land \Trans^*(\delta,s, \delta'',s') \limp
\delta'=\delta'',
\end{array}\]
\end{small}%
%
\noindent
For example, program $(a;b) \mid (a;c)$ is not SD, while
$a;(b\mid c)$ is (assuming the actions involved are always executable).
Thus, a (partial) execution of a SD
program is uniquely determined by the sequence of actions it has
produced. 
%


\section{Refinement Mappings} \label{sec:RefMap}

Suppose that we have a basic action theory $\D_l$ and another basic
action theory $\D_h$.  We would like to characterize whether $\D_h$ is
a reasonable abstraction of $\D_l$.  Here, we consider $\D_l$ as representing
the \emph{low-level} (LL) (or \emph{concrete}) action theory/agent and
$\D_h$ the \emph{high-level} (HL) (or \emph{abstract}) action
theory/agent.
%
%
We assume that $\mathit{\D_h}$ (resp. $\mathit{\D_l}$) involves a finite set of primitive action types
$\mathit{\A_h}$ (resp. $\mathit{\A_l}$) and a finite set of primitive fluent predicates $\mathit{\F_h}$ (resp. $\mathit{\F_l}$).
%
%
%
For simplicity, we assume that $\mathit{\D_h}$ and $\mathit{\D_l}$, share no domain
specific symbols except for the set of standard names for objects
$\mathit{\N}$.

We want to relate expressions in the language of $\D_h$ and
expressions in the language of $\D_l$.
We say that a function $m$ is a \emph{refinement mapping} from $\D_h$
to $\D_l$
if and only if:
\begin{enumerate}
\item for every high-level primitive action type $\mathit{A}$ in
  $\mathit{\A_h}$, $\mathit{m(A(\vec{x})) = \delta_A(\vec{x})}$, where
  $\delta_A(\vec{x})$ is a \ConGolog program over the low-level theory $\D_l$
  whose only free variables are $\vec{x}$, the parameters of the
  high-level action type; intuitively, $\delta_A(\vec{x})$ represents
  how the high-level action $A(\vec{x})$ can be implemented at the low
  level; 
since we use programs to specify the action sequences the agent may perform,
we require that $\delta_A(\vec{x})$ be situation-determined, i.e., the remaining program is always
    uniquely determined by the situation; 

  \item for every high-level primitive fluent 
    $F(\vec{x})$ (situation-suppressed) in $\mathit{\F_h}$, 
$\mathit{m(F(\vec{x})) = \phi_F(\vec{x})}$, where
$\phi_F(\vec{x})$ is a situation-suppressed formula over the language of
$\mathit{\D_l}$, and the only free variables are $\vec{x}$, the
object parameters of the high-level fluent; intuitively
$\phi_F(\vec{x})$ represents the \emph{low-level condition} under which
$F(\vec{x})$ holds in a situation.
\end{enumerate}

Note that we can map a fluent in the high-level theory to a fluent in
the low-level theory, i.e., 
$m(F_{h}(\vec{x})) = F_{l}(\vec{x})$,
which effectively amounts to having the low-level fluent be present in the high-level
theory. Similarly, one can include low-level actions in the high-level
theory.

\begin{figure}[htp]
	\centering
		\includegraphics[width=26mm,height=22mm]{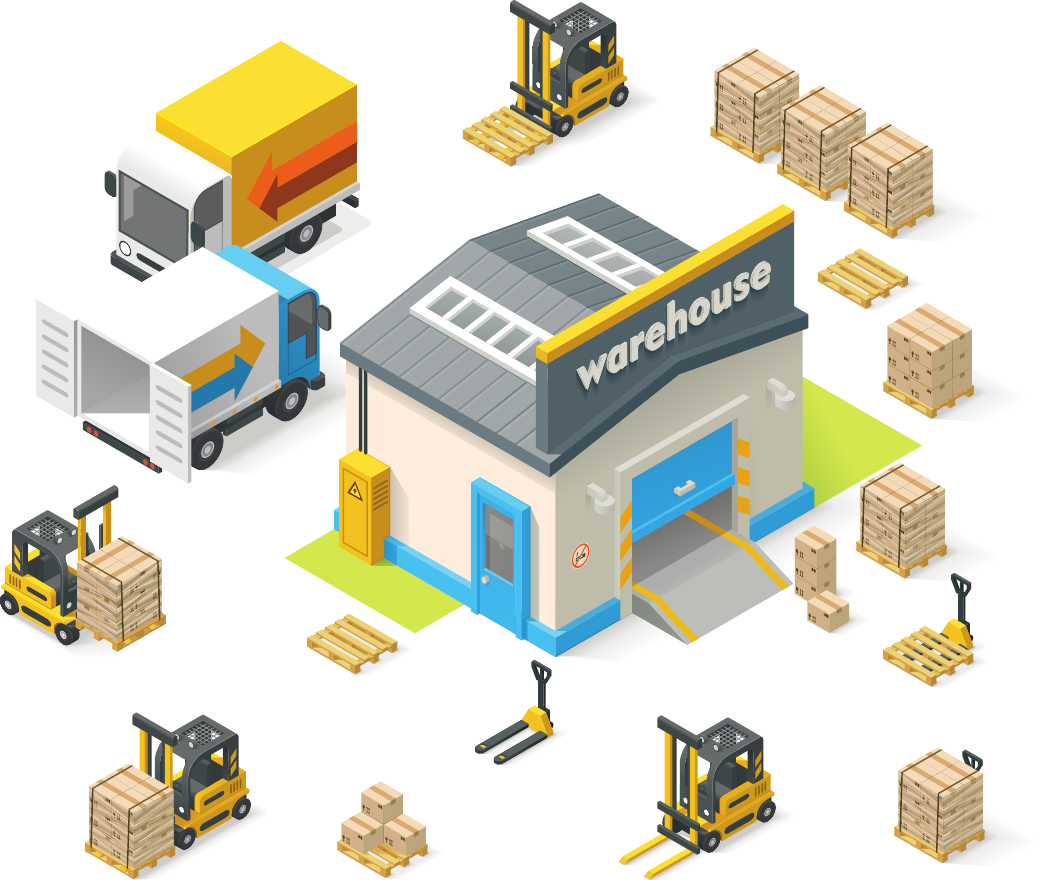}
		\includegraphics[width=44mm,height=24mm]{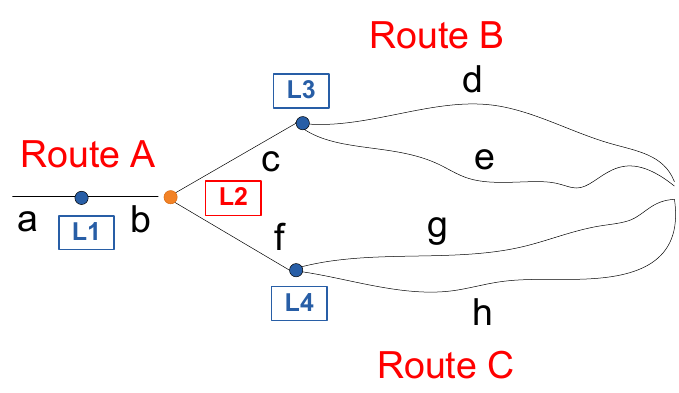}
		\includegraphics[width=24mm,height=21mm]{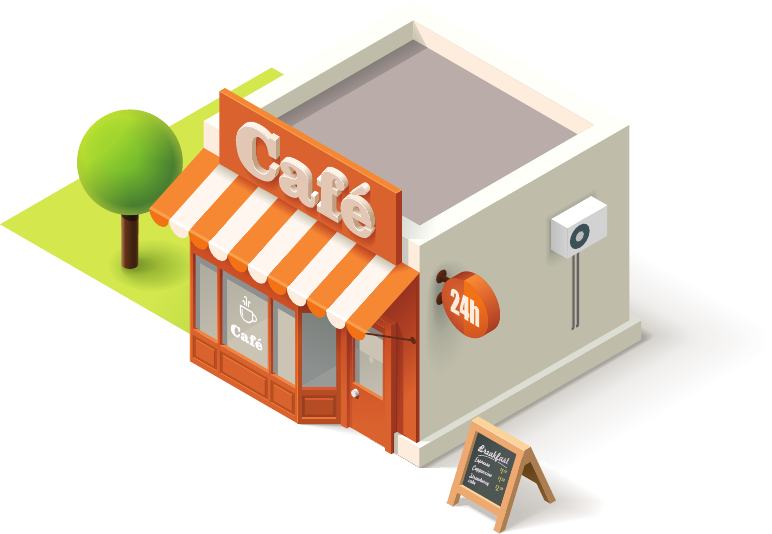}
	\caption{Transport Logistics Example}
	\label{fig:OfflineWCLocationsHLIn}
\end{figure}

We can extend the mapping to an arbitrary high-level situation-suppressed
formula $\phi$ by taking $m(\phi)$ to be the result of substituting
every fluent $F(\vec{x})$ in $\phi$ by  $m(F(\vec{x}))$.
Also, we can extend the mapping to sequences of high-level actions by
taking:
$m(\alpha_1,\ldots,\alpha_n) \doteq  m(\alpha_1);\ldots;m(\alpha_n)$ for $n
\geq 1$ and $m(\epsilon) \doteq nil$.
\\[1ex] 
\noindent
\textbf{Example.}
For our running example, we use a simple logistics domain.  There is 
a shipment with ID $123$ 
that is initially at a warehouse ($W$), and needs to be delivered to 
a cafe ($\mathit{Cf}$), along a network of roads shown in Figure \ref{fig:OfflineWCLocationsHLIn}.\footnote{Warehouse and cafe images are from freedesignfile.com.}

\paragraph{High-Level BAT $\D_h^{eg}$}
At the high level, we abstract over navigation and delivery procedure
details.  We have actions that represent choices of major routes
and delivering a shipment.
$\D_h^{eg}$ includes the following precondition axioms (throughout the
paper, we assume that free variables are universally quantified from
the outside):
%
\noindent
\begin{small}
\[\begin{array}{l} 
Poss(takeRoute(\mathit{sID}, r, o, d), s) \equiv  o \neq d \land At_{HL}(\mathit{sID}, o, s)  \land \\
\hspace*{5em} {}  
CnRoute_{HL}(r, o, d,s) 
\land (r=Rt_B \limp \neg Priority(\mathit{sID},s))
\\[0.5ex]
Poss(deliver(\mathit{sID}), s) \equiv \exists l.Dest_{HL}(\mathit{sID},l,s) 
{} \land At_{HL}(\mathit{sID}, l, s)
\end{array} \]
\end{small}%
\noindent
The action $takeRoute(\mathit{sID}, r, o,d)$ can be performed to take shipment
with ID $\mathit{sID}$ from origin location $o$ to destination location $d$ via
route $r$ 
(see Figure \ref{fig:OfflineWCLocationsHLIn}), and
is executable when the shipment is initially at $o$ and route $r$
connects $o$ to $d$; moreover,  priority shipments cannot be sent by
route $Rt_B$ (note that we refer to route X in Figure \ref{fig:OfflineWCLocationsHLIn} as $Rt_X$).  Action $deliver(\mathit{sID})$ can be performed to deliver
shipment $\mathit{sID}$ and is executable when $\mathit{sID}$ is at its destination.

The high-level BAT also includes the following 
SSAs:
\begin{small}
\[\begin{array}{l}
 At_{HL}(\mathit{sID}, l, do(a,s)) \equiv 
\exists l',r. a = takeRoute(\mathit{sID}, r, l', l) \; \lor \\
\hspace*{5em} At_{HL}(\mathit{sID}, l, s) \; \land \forall l', r. a \neq  takeRoute(\mathit{sID}, r, l, l')
\\[0.5ex]
Delivered(\mathit{sID}, do(a,s)) \equiv
a=deliver(\mathit{sID}) \lor Delivered(\mathit{sID}, s)
\end{array} \]
\end{small}%
\noindent
For the other fluents, we have SSAs specifying that they are unaffected
by any action.

$\D_h^{eg}$ also contains the following initial state axioms:  

\begin{small}
\[\begin{array}{l} 
Dest_{HL}(123,\mathit{Cf},S_0),~~
At_{HL}(123, W, S_0),\\
CnRoute_{HL}(Rt_A, W, L2, S_0),~~
CnRoute_{HL}(Rt_B, L2, \mathit{Cf}, S_0),\\
CnRoute_{HL}(Rt_C, L2, \mathit{Cf}, S_0)
\end{array} \]
\end{small}%

\noindent
Note that it is not known whether $123$ is a priority shipment.

\paragraph{Low Level BAT $\D_l^{eg}$}
At the low level, we model navigation and delivery in a more detailed way.
The agent has a more detailed map with more locations and roads
between them.  He also takes road closures into account.  Performing
delivery involves unloading the shipment and getting a signature.
The low-level BAT $\D_l^{eg}$ includes the following action precondition axioms:
%
\noindent
\begin{small}
\[\begin{array}{l}
Poss(takeRoad(\mathit{sID}, t, o, d), s) \equiv  o \neq d {} \; \land  \\
\hspace*{1.5em} At_{LL}(\mathit{sID}, o, s) \land CnRoad(t, o, d,s) \land \neg Closed(t,s) \;  \land\\
\hspace*{1.5em} (d = L3 \limp \neg (BadWeather(s) \lor Express(\mathit{sID},s)))
\\[0.5ex]
Poss(unload(\mathit{sID}), s) \equiv \exists l.Dest_{LL}(\mathit{sID},l,s) \; \land
    At_{LL}(\mathit{sID},l, s)
\\[0.5ex]
Poss(getSignature(\mathit{sID}), s) \equiv Unloaded(\mathit{sID},s) 
\end{array} \]
\end{small}%
\noindent
Thus, the action $takeRoad(\mathit{sID}, t, o, d)$, where the agent takes
shipment $\mathit{sID}$ from origin location $o$ to destination $d$ via road
$t$, is executable provided that $t$ connects $o$ to $d$, $\mathit{sID}$ is at
$o$, and $t$ is not closed; moreover, a road to
$L3$ cannot be taken if the weather is bad or $\mathit{sID}$ is an express shipment as this would
likely violate quality of service requirements.


The low-level BAT includes the following SSAs: 

\begin{small}
\[\begin{array}{l}
Unloaded(\mathit{sID}, do(a,s)) \equiv 
a=unload(\mathit{sID}) \lor Unloaded(\mathit{sID}, s)
\\[0.5ex]
Signed(\mathit{sID}, do(a,s)) \equiv 
a=getSignature(\mathit{sID}) \lor Signed(\mathit{sID}, s)
\end{array} \]
\end{small}%

\noindent 
The SSA for $At_{LL}$ is like the one for $At_{HL}$ with
$takeRoute$ replaced by $takeRoad$.  For the other fluents, we have
SSAs specifying that they are unaffected by any actions. Note that we
could easily include exogenous actions for road closures and change in
weather, new shipment orders, etc.


$\D_l^{eg}$ also contains the following initial state axioms:

\noindent
\begin{small}
\[\begin{array}{l}
\neg BadWeather(S_0),~ Closed(r,S_0) \equiv r = Rd_e, \\
 Express(123,S_0),~ Dest_{LL}(123,\mathit{Cf},S_0),~ At_{LL}(123, W, S_0) 
 \end{array} \]
\end{small}%

\noindent
together with a complete specification of $CnRoad$ and $CnRoute_{LL}$
as in Figure \ref{fig:OfflineWCLocationsHLIn}. We refer to road x in
the figure as $Rd_x$.

\paragraph{Refinement Mapping $m^{eg}$}
We specify the relationship between the high-level and low-level BATs
through the following refinement mapping $m^{eg}$:

\begin{small}
\[\begin{array}{l}
m^{eg}(takeRoute(\mathit{sID}, r, o, d))= \\
\hspace{1.3em} (r = Rt_A \land CnRoute_{LL}(Rt_A,o,d))?; \\
\hspace{2em} \pi t. takeRoad(\mathit{sID}, t, o, L1); \pi t'. takeRoad(\mathit{sID}, t', L1,
    d) \mid {}
\\
\hspace{1.3em} (r = Rt_B \land CnRoute_{LL}(Rt_B,o,d))?; \\
\hspace{2em} \pi t. takeRoad(\mathit{sID}, t, o, L3 ); 
\pi t'. takeRoad(\mathit{sID}, t', L3, d ) \mid {}
\\
\hspace{1.3em} (r = Rt_C \land CnRoute_{LL}(Rt_C,o,d))?; \\
\hspace{2em} \pi t. takeRoad(\mathit{sID}, t, o, L4 ); 
\pi t'. takeRoad(\mathit{sID},t', L4, d)
\\[1ex]
m^{eg}(deliver(\mathit{sID})) = unload(\mathit{sID}); getSignature(\mathit{sID}) 
\\[0.5ex]
m^{eg}(Priority(\mathit{sID}))= BadWeather \lor Express(\mathit{sID})
\\[0.5ex]
m^{eg}(Delivered(\mathit{sID})) =  Unloaded(\mathit{sID}) \land  Signed(\mathit{sID})
\\[0.5ex]
m^{eg}(At_{HL}(\mathit{sID}, loc)) = At_{LL}(\mathit{sID}, loc)
\\[0.5ex]
m^{eg}(CnRoute_{HL}(r,o,d)) =  CnRoute_{LL}(r,o,d)
\\[0.5ex]
m^{eg}(Dest_{HL}(\mathit{sID},l)) =  Dest_{LL}(\mathit{sID},l)
\end{array} \]
\end{small}%

\noindent
Thus, taking route $Rt_A$ involves first taking a road from the origin $o$
to $L1$ and then taking another road from $L1$  to the destination $d$.
For the other two routes, the refinement mapping is similar except a different
intermediate location must be reached.
Note that we could easily write programs to specify refinements for more complex
routes, e.g., that take a sequence of roads from $o$ to $d$ going through 
intermediate locations belonging to a given set.
We refine the high-level fluent $Priority(\mathit{sID})$ to the condition where
either the weather is bad or the shipment is express.
$\qedf$


\section{$m$-Bisimulation} \label{sec:Bisim}

To relate high-level and low-level models/theories,
we resort to a suitable notion of bisimulation, i.e., one that is
relative to the refinement mapping.
Let $M_h$ be a model of the high-level BAT $\D_h$, $M_l$ a model
of the low-level BAT $\D_l \cup \C$, and $m$ a refinement mapping from
$\D_h$ to $\D_l$.

We first define a local condition for the bisimulation.
We say that situation $s_h$ in $M_h$ is $m$-isomorphic to situation
$s_l$ in $M_l$, written $s_h \simeq_m^{M_h,M_l} s_l$, if and only if
\[M_h,v[s/s_h] \models F(\vec{x},s) \mbox{ if and only if } M_l,v[s/s_l] \models
  m(F(\vec{x}))[s]\]
 for every high-level
  primitive fluent $F(\vec{x})$ in $\mathit{\F_h}$ and every variable assignment
  $v$  ($v[x/e]$ 
	stands for the assignment that is like $v$
  except that $x$ is mapped to $e$), 
  i.e., $s_h$ and $s_l$ interpret all high-level fluents the same. 


A relation $B \subseteq \Delta_S^{M_h} \times \Delta_S^{M_l}$ 
(where $\Delta_S^{M}$ stands for the situation domain of $M$)  is an
\emph{$m$-bisimulation relation between $M_h$ and $M_l$}  
if $\langle s_h,
s_l \rangle \in B$ implies that:
\begin{enumerate}
\item $s_h \simeq_m^{M_h,M_l} s_l$, i.e., $s_h$ in $M_h$ is $m$-isomorphic to situation
$s_l$ in $M_l$;

\item
for every high-level primitive action type $\mathit{A}$ in
$\mathit{\A_h}$, 
if  there exists $s_h'$ such that 
$M_h,v[s/s_h,s'/s_h'] \models  Poss(A(\vec{x}),s) \land s' = do(A(\vec{x}),s)$, 
then there exists  $s_l'$ such that
$M_l,v[s/s_l,s'/s_l'] \models Do(m(A(\vec{x})),s,s')$ and
$\langle s_h',s_l' \rangle \in B$,
i.e., if $A(\vec{x})$ is executable in the high-level model at $s_h$, then the program that
implements $A(\vec{x})$ must be executable in the low-level model at $s_l$, and the resulting
pair of situations $s_h'$ and $s_l'$ must be bisimilar;

\item for every high-level primitive action type $\mathit{A}$ in
  $\mathit{\A_h}$, if there exists  $s_l'$ such that
  $M_l,v[s/s_l,s'/s_l'] \models Do(m(A(\vec{x})),s,s')$, then
  there exists  $s_h'$ such that
  $M_h,v[s/s_h,s'/s_h'] \models Poss(A(\vec{x}),s) \land s' =
  do(A(\vec{x}),s)$ and $\langle s_h',s_l' \rangle \in B$,
  i.e., if the program that implements $A(\vec{x})$ is executable in the low-level model at $s_l$, 
	then $A(\vec{x})$ must be executable in the high-level model at $s_h$, and the resulting
  pair of situations $s_h'$ and $s_l'$ must be bisimilar.
\end{enumerate}

We say that \emph{$M_h$ is bisimilar to $M_l$ relative to refinement mapping
$m$, i.e., $m$-bisimilar, written $M_h \sim_m M_l$}, if and only if
there exists an $m$-bisimulation relation $B$ between  $M_h$ and $M_l$ such
that $\langle S_0^{M_h},  S_0^{M_l} \rangle \in B$.
A situation $s_h$ in $M_h$ is $m$-bisimilar to situation $s_l$ in $M_l$, 
written $s_h \sim_m^{M_h,M_l} s_l$, if and only if
there exists an $m$-bisimulation relation $B$ between  $M_h$ and $M_l$ 
and $\langle s_h, s_l \rangle \in B$.


Given these definitions, we immediately get the following results.
First, we can show that $m$-isomorphic situations
satisfy the same high-level situation-suppressed formulas:\footnote{For proofs of all our results, see the appendix.}
\begin{lem} \label{lem:sitSupBisim} 
 If $s_h \simeq_m^{M_h,M_l} s_l$, then for any high-level
  situation-suppressed formula $\phi$, we have that:
\\[0.5ex]  
\begin{small}  
\[\begin{array}{l}
\hspace*{1em} M_h,v[s/s_h] \models \phi[s]\ \ \mbox{if and only if}\ \ 
M_l,v[s/s_l] \models m(\phi)[s].
\end{array}\]
\end{small}%
\end{lem}





We can also show that in $m$-bisimilar
models, the same sequences of high-level actions are executable, and
that the resulting situations are $m$-bisimilar:

\begin{lem} \label{lem:bisimL2HOffND}
If  $M_h \sim_m M_l$, then
for any sequence of high-level actions $\vec{\alpha}$,
we have that

\begin{small}  
\[\begin{array}{l}
\mbox{if } M_l,v[s'/s_l] \models Do(m(\vec{\alpha}),S_0,s'), 
    \mbox{then there exists $s_h$ such that}\\
    M_h,v[s'/s_h]  \models s' = do(\vec{\alpha},S_0) \land
    Executable(s') \mbox{ and } s_h \sim_m^{M_h,M_l} s_l\\[1ex]
    \mbox{and } \\[1ex]
\mbox{if }  M_h,v[s'/s_h]  \models s_h = do(\vec{\alpha},S_0) \land
    Executable(s_h),\\
\mbox{then there exists $s_l$ such that } M_l,v[s'/s_l] \models Do(m(\vec{\alpha}),S_0,s')
 \mbox{ and } s_h \sim_m^{M_h,M_l} s_l.
\end{array}\]
\end{small}%
\end{lem}


Given the above results, it is straightforward to show that in $m$-bisimilar
models, the same sequences of high-level actions are executable, and
in the resulting situations, the same high-level
situation-suppressed formulas hold:

\begin{thm} \label{thm:bisimL2HOff}
  If  $M_h \sim_m M_l$, then 
  for any sequence of ground high-level actions $\vec{\alpha}$ and
  any high-level situation-suppressed formula $\phi$, 
  we have that

\begin{small}  
\[\begin{array}{l}
  M_l \models \exists s' Do(m(\vec{\alpha}),S_0,s') \land
  m(\phi)[s'] \hspace{1em} \mbox{ if and only if }\\
  \hspace{4em} M_h \models Executable(do(\vec{\alpha},S_0)) \land
  \phi[do(\vec{\alpha},S_0)].
\end{array}\]
\end{small}%

\end{thm}



\section{Sound Abstraction} \label{sec:Sound}

To ensure that the high-level theory is consistent with the low-level
theory and mapping $m$, we may require that for every model of the low-level
theory, there is an $m$-bisimilar structure that is a model of the
high-level theory.

We say that $\D_h$ is a
\emph{sound abstraction of} $\D_l$ \emph{relative to refinement
  mapping} $m$ if and only if, for every model $M_l$ of $\D_l \cup \C$, there
exists a model $M_h$ of $\D_h$ such that  $M_h \sim_m M_l$.
\\[1ex]
\noindent
\textbf{Example Cont.}
Returning to our example of Section~ \ref{sec:RefMap}, it is straightforward to
show that it involves a high-level theory $\D_h^{eg}$ that is a sound abstraction of
the low-level theory $\D_l^{eg}$ relative to the mapping $m^{eg}$.
We discuss how we prove this later. $\qedf$

Sound abstractions have many interesting and useful properties.
First, from the definition of sound abstraction and Theorem \ref{thm:bisimL2HOff},
we immediately get the following result:
\begin{corollary} \label{cor:ExecutableL2HOffSat}
Suppose that $\D_h$ is a sound abstraction of $\D_l$ relative to 
mapping $m$. Then for any sequence of ground high-level actions $\vec{\alpha}$ and for
  any high-level situation-suppressed formula $\phi$,
if $\D_l \cup \C \cup \{ \exists s.
Do(m(\vec{\alpha}),S_0,s) \land m(\phi)[s]\}$ is satisfiable, then 
$\D_h \cup
\{ Executable(do(\vec{\alpha},S_0)) \land \phi[do(\vec{\alpha},S_0)]
\}$ is also satisfiable.  
In particular, if $\D_l \cup \C \cup \{ \exists s.
Do(m(\vec{\alpha}),S_0,s) \}$ is satisfiable, then 
$\D_h \cup
\{ Executable(do(\vec{\alpha},S_0))
\}$ is also satisfiable.
\end{corollary}
Thus if the low-level agent/theory thinks that a refinement of
$\vec{\alpha}$ (perhaps involving exogenous actions) may occur (with
$m(\phi)$ holding afterwards), the high-level
agent/theory also thinks that $\vec{\alpha}$ may occur
(with $\phi$ holding afterwards).  
If we observe that such a refinement actually occurs it will
thus be consistent with the high-level theory.

We can also show that if the high-level theory entails that some
sequence of high-level actions $\vec{\alpha}$ is executable, and that in the resulting
situation, a situation-suppressed formula $\phi$  holds, then the
low-level theory must also entail that some refinement of
$\vec{\alpha}$ is executable and that in the resulting 
situation $m(\phi)$  holds:

\begin{thm} \label{Thm:ExecutableL2HOffEntail}
Suppose that $\D_h$ is a sound abstraction of $\D_l$ relative to 
mapping $m$.  Then for any ground high-level action sequence $\vec{\alpha}$ and
for any high-level situation-suppressed formula $\phi$,
if $\D_h \models Executable(do(\vec{\alpha},S_0)) \land \phi[do(\vec{\alpha},S_0)]$, then
$\D_l \cup \C \models \exists s. Do(m(\vec{\alpha}),S_0,s) \land m(\phi)[s]$.
\end{thm}


We can immediately relate the above result to \emph{planning}.
In the situation calculus, the planning problem is usually defined as
follows \cite{Reiter01-Book}:
\begin{quote}
Given a BAT $\D$, and a situation-suppressed goal formula $\phi$, 
find a ground action sequence $\vec{a}$ such that
$\D \models Executable(do(\vec{a},S_0)) \land \phi[do(\vec{a},S_0)]$.
\end{quote}
%
Thus, Theorem \ref{Thm:ExecutableL2HOffEntail} means that if we can
find a plan $\vec{\alpha}$ to achieve a goal $\phi$ at the high level,
i.e.,
$\D_h \models Executable(do(\vec{\alpha},S_0)) \land
\phi[do(\vec{\alpha},S_0)]$,
then it follows that there exists a refinement of $\vec{\alpha}$ that
achieves $\phi$ at the low level, i.e.,
$\D_l \cup \C \models \exists s. Do(m(\vec{\alpha}),S_0,s) \land
m(\phi)[s]$.
However, note that the refinement could in general be different from model
to model.  But if, in addition, we have complete information at
the low level, i.e., a single model for $\D_l$, then, since we have
standard names and domain closure for objects and actions, we can always obtain a plan to
achieve the goal $\phi$ by finding a refinement in this way, i.e.,
there exists a ground low-level action sequence $\vec{a}$ such that 
$\D_l \cup \C \models Do(m(\vec{\alpha}),S_0,do(\vec{a},S_0)) \land
m(\phi)[do(\vec{a},S_0)]$.
The search space of refinements of $\vec{\alpha}$ would typically be
much smaller than the space of all low-level action sequences,
thus yielding important efficiency benefits.


We can also show that if $\D_h$ is a sound abstraction of $\D_l$ with
respect to a mapping, 
then the different sequences of low-level actions that are refinements
of a given high-level primitive action sequence all have the same
effects on the high-level fluents, and more generally on high-level
situation-suppressed formulas, i.e., from the high-level
perspective they are deterministic:
\begin{corollary} \label{cor:HLdeterminismOfLL}
If  $\D_h$ is a sound abstraction of $\D_l$ relative to
mapping $m$, then for any sequence of ground high-level actions
$\vec{\alpha}$ and for any high-level situation-suppressed formula
$\phi$, we have that

\begin{small}
\[\begin{array}{l}
\D_l \cup \C \models \forall s \forall s'. Do(m(\vec{\alpha}),S_0,s) \land
Do(m(\vec{\alpha}),S_0,s') \limp
    (m(\phi)[s] \equiv m(\phi)[s'])
\end{array}\]
\end{small}%
\end{corollary}

\noindent
An immediate consequence of the above is the following:
\begin{corollary} \label{cor:HLdeterminism2OfLL}
If  $\D_h$ is a sound abstraction of $\D_l$ relative to
mapping $m$, then for any sequence of ground high-level actions
$\vec{\alpha}$ and for any high-level situation-suppressed formula
$\phi$, we have that

\begin{small}
\[\begin{array}{l}
\D_l \cup \C \models 
(\exists s. Do(m(\vec{\alpha}),S_0,s) \land m(\phi)[s]) \supset
    (\forall s. Do(m(\vec{\alpha}),S_0,s) \limp m(\phi)[s])
\end{array}\]
\end{small}%
\end{corollary}

It is also easy to show that if some refinement of the 
sequence of high-level actions $\vec{\alpha}\beta$ is executable, then 
there exists a refinement of $\beta$ that is executable after
executing any refinement of $\vec{\alpha}$:
\begin{thm} \label{thm:ExecutableL2HOffEntail2}
If  $\D_h$ is a sound abstraction of $\D_l$ relative to
mapping $m$, then for any sequence of ground high-level actions
$\vec{\alpha}$ and for any ground high-level action $\beta$, we have that

\begin{small}
\[\begin{array}{l}
\D_l \cup \C \models 
\exists s. Do(m(\vec{\alpha}\beta),S_0,s) \supset
    (\forall s. Do(m(\vec{\alpha}),S_0,s) \limp \exists s'. Do(m(\beta),s,s'))
\end{array}\]
\end{small}%
\end{thm}

\noindent
Notice that this applies to all prefixes of $\vec{\alpha}$, so using
Corollary \ref{cor:HLdeterminism2OfLL} as well, we immediately get that:
\begin{corollary} \label{cor:ExecutableL2HOffEntail3}
Suppose that $\D_h$ is a sound abstraction of $\D_l$ relative to 
mapping $m$.  Then for any ground high-level action sequence
$\alpha_1,\ldots,\alpha_n$, and
for any high-level situation-suppressed formula $\phi$, then we have that:

\begin{small}
\[\begin{array}{l}
\D_l \cup \C \models 
(\exists s. Do(m(\alpha_1,\ldots,\alpha_n),S_0,s) \land  m(\phi)[s]) \supset\\
\hspace{4.2em}((\forall s. Do(m(\alpha_1,\ldots,\alpha_n),S_0,s)
    \supset m(\phi)[s]) \land {}\\
\hspace{4.5em} (\exists s. Do(m(\alpha_1),S_0,s)) \land {}\\
\hspace{4.5em} \bigwedge_{2 \leq i \leq n} (\forall
    s. Do(m(\alpha_1,\ldots,\alpha_{i-1}),S_0,s) \supset \\
\hspace{9.4em} \exists s'. Do(m(\alpha_i),s,s')))
\end{array}\]
\end{small}%
\end{corollary}

These results mean that if a ground high-level action sequence achieves a
high-level condition $\phi$, we can choose refinements of the actions in the
sequence independently and be certain to obtain a refinement of the
complete sequence that achieves $\phi$.
We can exploit this in planning to gain even more efficiency.  If we
can find a plan $\alpha_1, \ldots,\alpha_n$ to achieve a goal $\phi$
at the high level, then there exists a refinement of
$\alpha_1, \ldots,\alpha_n$ that achieves $m(\phi)$ at the low level, and
we can obtain it by finding refinements of the high-level actions
$\alpha_i$ for $i: 1 \leq i \leq n$ one by one, without ever having to
backtrack.  The search space would typically be exponentially smaller in the
length of the high-level plan $n$.  If we have complete information at
the low level, then we can always obtain a refined plan to achieve
$m(\phi)$ in this way.
\\[1ex]
\noindent
\textbf{Example Cont.} 
In our running example
we can show that the action sequence 
$\vec{\alpha}=$ 
$[takeRoute(123, Rt_A, W, L2),$ $takeRoute(123, Rt_C, L2,\mathit{Cf}),$ \linebreak $deliver(123)]$
is a valid high-level plan to achieve the goal $\phi_g = Delivered(123)$ of having delivered
shipment $123$, i.e.,
$\D_h^{eg} \models Executable(do(\vec{\alpha},S_0)) \; \land $ \linebreak $\phi_g[do(\vec{\alpha},S_0)].$
Since $\D_h^{eg}$ is a sound abstraction of the low-level theory
$\D_l^{eg}$ relative to the mapping $m^{eg}$, we know that we can find
a refinement of the high-level plan $\vec{\alpha}$ that achieves the
refinement of the goal
$m^{eg}(\phi_g) = Unloaded(123) \land Signed(123)$.
In fact, we can show that
$
\D_l^{eg} \cup \C \models Do(m^{eg}(\vec{\alpha}),$\linebreak$S_0,do(\vec{a}\vec{b}
\vec{c},S_0)) \land m^{eg}(\phi_g)[do(\vec{a}\vec{b}\vec{c},S_0)]
$
for $\vec{a} = [takeRoad(123, Rd_a, W, L1), $ \linebreak $takeRoad(123, Rd_b,L1, L2)]$,
$\vec{b} = [takeRoad(123, Rd_f, L2, L4),$ $takeRoad(123, $ $ Rd_g, L4, \mathit{Cf})]$, and 
$\vec{c} = [unload(123),$ $getSignature(123)]$.
$\qedf$

Now, let us define some low-level programs that characterize
the refinements of high-level action/action sequences:

\begin{small}
\[\begin{array}{l}
\mathit{\anyonehl} \doteq \ \ \mid_{A_i \in \A_h} \pi
    \vec{x}.m(A_i(\vec{x}))
    \\ 
    \qquad ~\mbox{i.e., do any refinement of any one HL primitive action,}
\\[1ex]
\mathit{\anyseqhl} \doteq \mathit{\anyonehl}^*
\\ 
\qquad ~~\mbox{i.e., do any sequence of refinements of HL actions.}
\end{array}\]
\end{small}%

How does one verify that one has a sound abstraction?  The following
result identifies necessary and sufficient conditions for having a sound
abstraction:
\begin{thm} \label{thm:verifySound} 
 $\D^h$ is a sound abstraction of $\D^l$ relative to mapping $m$ if and only if
\begin{small}
\begin{description}
\item[(a)]
$\D^l_{S_0} \cup \D^l_{ca} \cup \D^l_{coa} \models m(\phi)$, for all
$\phi \in D^h_{S_0}$,
\item[(b)]
$\mathit{\D^l \cup \C \models \forall s. Do(\anyseqhl,S_0,s) \limp}$ \\
\hspace*{1em} $\bigwedge_{A_i \in \A^h}
    \forall \vec{x}.
 (m(\phi^{Poss}_{A_i}(\vec{x}))[s] 
    \equiv \exists s' Do(m(A_i(\vec{x})),s,s')),$
\item[(c)]
$\D^l \cup \C \models \forall s. Do(\anyseqhl,S_0,s) \limp $ \\
\hspace*{2em} $\bigwedge_{A_i \in \A^h}  \forall \vec{x}, s'. (Do(m(A_i(\vec{x})),s,s') \limp$ \\
\hspace*{4.1em} $\bigwedge_{F_i \in \F^h}  \forall \vec{y}  (m(\phi^{ssa}_{F_i,A_i}(\vec{y},\vec{x}))[s] \equiv m(F_i(\vec{y}))[s'])),$
\end{description}
\end{small}%
\noindent
where $\phi^{Poss}_{A_i}(\vec{x})$ is the right hand side (RHS) of the precondition axiom for action $A_i(\vec{x})$, and $\phi^{ssa}_{F_i,A_i}(\vec{y},\vec{x})$ is the RHS of the successor state axiom for $F_i$ instantiated with action
$A_i(\vec{x})$ where action terms have been eliminated using
$\D^h_{ca}$. 
\end{thm}

\noindent
The above 
provides us with a way of showing
that we have a sound abstraction by proving that certain properties are entailed by the low-level theory. 
Condition (a) is straightforward to verify and 
conditions (b) and (c) are properties of programs
that standard verification techniques can deal with.
The theorem also means that if $\D^h$ is a sound abstraction of $\D^l$ with respect to $m$, 
then $\D^l$ must entail the mapped high-level successor state axioms
and entail that the mapped conditions for a high-level action to be executable
(from the precondition axioms of $\D^h$) correctly capture the
executability conditions of their refinements (these conditions must
hold after any sequence of refinements of high-level actions, i.e.,
in any situation $s$ where $Do(\anyseqhl,S_0,s)$ holds).
\\[1ex]

Returning to our running example,
it is straightforward to
show that it involves a high-level theory $\D_h^{eg}$ that is a sound abstraction of
the low-level theory $\D_l^{eg}$ relative to the mapping $m^{eg}$:

\begin{proposition} \label{prop:DhEgSoundAbsElEg}
$\D^{eg}_h$ is a sound abstraction of $\D^{eg}_l$ wrt $m^{eg}$.
\end{proposition}

\noindent
$\D_l^{S_0}$ entails the ``translation'' of all the facts about the high-level
fluents $CnRoute_{HL}$, $Dest_{HL}$ and $At_{HL}$ that are
in $\D_h^{S_0}$. 
For instance, in the high level theory, we have that $\D_h^{eg} \models Dest_{HL}(123,\mathit{Cf},S_0)$. We have that $Dest_{HL}(\mathit{sID})$ is mapped to $Dest_{LL}(\mathit{sID},l)$ and furthermore, in the low-level theory, we have that $\D_l^{eg} \models Dest_{LL}(123,\mathit{Cf},S_0)$.  
Moreover, $\D_l^{eg}$ entails that the mapping of the
preconditions of the high-level actions $deliver$ and $takeRoute$
correctly capture the executability conditions of their
refinements. 
For example, high-level action $deliver(\mathit{sID})$ is mapped to
the program $unload(\mathit{sID});$\linebreak $getSignature(\mathit{sID})$ in the low-level theory. Also, the precondition of\linebreak  $deliver(\mathit{sID})$ 
maps to $\exists l.Dest_{LL}(\mathit{sID},l,s) \land At_{LL}(\mathit{sID}, l, s)$ which is the precondition for of $unload(\mathit{sID})$ and 
this in turn ensures the precondition for action $getSignature(\mathit{sID})$.
$\D_l^{eg}$ also entails that the mapped high-level successor
state axioms hold at the low level for refinements of high-level actions.
For instance, consider the high-level fluent $Delivered(\mathit{sID},s')$, which is only affected by action $deliver(\mathit{sID})$. 
Given that $deliver(\mathit{sID})$ is mapped to the program $unload(\mathit{sID});getSignature(\mathit{sID})$ 
and $Delivered(\mathit{sID})$ maps to $Unloaded(\mathit{sID}) \land Signed(\mathit{sID})$ in the low-level theory, we can use
the successor state axioms of $Unloaded(\mathit{sID},s')$ and
$Signed(\mathit{sID},s')$ to show that the result holds.
For high-level fluent $At_{HL}$, which is only affected by the
$takeRoute$ action, the proof is similar.
Thus, $\D_h^{eg}$ is a sound
abstraction of $\D_l^{eg}$ relative to $m^{eg}$.
%
For more details, see the proof in the appendix.
$\qedf$



\section{Complete Abstraction} \label{sec:Complete}

When we have a sound abstraction $\D_h$ of a low-level theory $\D_l$
with respect to a mapping $m$, the high-level theory $\D_h$'s conclusions are
always sound with respect to the more refined theory $\D_l$, but $\D_h$ may have
less information than $\D_l$ regarding high-level actions and
conditions. $\D_h$ may consider it possible that a high-level action sequence 
is executable (and achieves a goal) when $\D_l$ knows it is not.
The low-level theory may know/entail that a refinement of a high-level action sequence 
achieves a goal without the high level knowing/entailing it. 
We can define a stronger notion of abstraction that ensures that the
high-level theory knows everything that the low-level theory knows
about high-level actions and conditions.

We say that $\D_h$ is a
\emph{complete abstraction of} $\D_l$ \emph{relative to refinement
  mapping} $m$ if and only if, for every model $M_h$ of $\D_h$, there
exists a model $M_l$ of $\D_l \cup \C$ such that  $M_l \sim_m M_h$.

From the definition of complete abstraction and Theorem
\ref{thm:bisimL2HOff}, we immediately get the following converses of Corollary \ref{cor:ExecutableL2HOffSat}
and Theorem \ref{Thm:ExecutableL2HOffEntail}:
\begin{corollary} \label{cor:ExecutableL2HOffSatComp}
Suppose that $\D_h$ is a complete abstraction of $\D_l$ relative to
$m$. Then for any sequence of ground high-level actions $\vec{\alpha}$ and for
  any high-level situation-suppressed formula $\phi$,
if  
$\D_h \cup
\{ Executable(do(\vec{\alpha},S_0)) \land \phi[do(\vec{\alpha},S_0)]
\}$ is satisfiable, 
then
$\D_l \cup \C \cup \{ \exists s.
Do(m(\vec{\alpha}),S_0,s) \land m(\phi)[s]\}$ is satisfiable.
In particular, if $\D_h \cup
\{ Executable(do(\vec{\alpha},S_0))
\}$ is satisfiable, then $\D_l \cup \C \cup \{ \exists s.
Do(m(\vec{\alpha}), $ $ S_0,s) \}$ is satisfiable.
\end{corollary}

\begin{thm} \label{Thm:ExecutableL2HOffEntailComp}
Suppose that $\D_h$ is a complete abstraction of $\D_l$ relative to
mapping $m$.  Then for any ground high-level action sequence $\vec{\alpha}$ and
any high-level situation-suppressed formula $\phi$, if
$\D_l \cup \C \models \exists s. Do(m(\vec{\alpha}),S_0,s) \land
m(\phi)[s]$,
then
$\D_h \models Executable(do(\vec{\alpha},S_0)) \land \phi[do(\vec{\alpha},S_0)]$.
\end{thm}

Thus when we have a high-level theory $\D_h$ that is a complete
abstraction of a low-level theory $\D_l $ with respect to a mapping $m$, if
$\D_l$ knows/entails that some refinement of a high-level action sequence $\vec{\alpha}$
achieves a high-level goal $\phi$, then $\D_h$ knows/entails that $\vec{\alpha}$
achieves $\phi$.  It follows that we can find all high-level plans to achieve
high-level goals using $\D_h$.

Note as well that with a complete abstraction that is not a sound
abstraction, we no longer get that high-level actions are deterministic at
the low level with respect to high-level fluents, i.e., Corollary
\ref{cor:HLdeterminismOfLL}; this happens because $\D_l \cup \C$ may
have models that are not $m$-bisimilar to any model of $\D_h$ and where different
refinements of a high-level action yield different truth-values for
$m(F)$, for some high-level fluent $F$.

Complete abstractions can constrain the search space that is used in
automated reasoning (e.g., planning) and thus speed up finding solutions to problems.
But if we don't have soundness, we need to check that the result
actually holds at the low level.

We also say that $\D_h$ is a \emph{sound and complete abstraction of}
  $\D_l$ \emph{relative to refinement mapping} $m$ if and only if
  $\D_h$ is both a sound and a complete abstraction of $\D_l$ relative
  to $m$.
\\[1ex]
\noindent
\textbf{Example Cont.}~Returning to our running example, the high-level
theory does not know whether shipment $123$ is high
priority, i.e., \linebreak
$\D_h^{eg} \not\models Priority(123)[S_0]$ and
$\D_h^{eg} \not\models \lnot Priority(123)[S_0]$, but the low-level
theory knows that it is, i.e.,
$\D_l^{eg} \models m^{eg}(Priority(123))[S_0]$.  Thus $\D_h^{eg}$ has a
model where $\lnot Priority(123)[S_0]$ holds that is not
$m^{eg}$-bisimilar to any model of $\D_l^{eg}$, and thus $\D_h^{eg}$ is a
sound abstraction of $\D_l^{eg}$ with respect to $m^{eg}$, but not a
complete abstraction.  For instance, the high-level theory considers
it possible that the shipment can be delivered by taking route A and
then route B, i.e.,
$\D_h^{eg} \cup \{Executable(do(\vec{\alpha},S_0)) \land
\phi_g[do(\vec{\alpha},S_0)]\}$
is satisfiable for
$\vec{\alpha}= [takeRoute(123,Rt_A,W,L2), takeRoute(123,Rt_B,$
$L2,\mathit{Cf}), deliver(123)]$
and $\phi_g =$ $Delivered(123)$.  But the low-level theory knows that
$\vec{\alpha}$  cannot be refined to an executable low-level plan, i.e.,
$\D_l^{eg} \cup \C \models \lnot \exists s. Do(m^{eg}(\vec{\alpha}),S_0,s)$.
If we add $Priority(123)[S_0]$ and a complete specification of $CnRoute_{HL}$ to 
$\D_h^{eg}$,
then it becomes a sound and complete abstraction of $\D_l^{eg}$ with respect to
$m^{eg}$.
The plan $\vec{\alpha}$ is now ruled out as 
$\D_h^{eg} \cup \{ Priority(123,$ $ S_0)  \} \cup \{Executable(do(\vec{\alpha},S_0)) \}$
is not satisfiable.
$\qedf$

The following result provides a method to verify that
we have a sound and complete abstraction:
\begin{thm} \label{thm:verifySoundComplete}
If $\D^h$ is a sound abstraction of $\D^l$ relative to mapping $m$, then
$\D^h$ is also a complete abstraction of $\D^l$ with respect to mapping $m$
if and only if
for every model $M_h$ of $\D^h_{S_0} \cup \D^h_{ca} \cup \D^h_{coa}$,
there exists a model  $M_l$ of $\D^l_{S_0} \cup \D^l_{ca}
\cup\D^l_{coa}$ such that $S_0^{M_h} \simeq_m^{M_h,M_l} S_0^{M_l}$.
\end{thm}
We also have the following result that characterizes complete (but not
necessarily sound) abstractions:
\begin{thm} \label{thm:verifyComplete}
$\D^h$ is a complete abstraction of $\D^l$ relative to mapping $m$ if
and only if
  for every model $M_h$ of $\D^h$,
  there exists a model  $M_l$ of $\D^l \cup \C$ such that

\begin{small}
\begin{description}
\item[(a)]
  $S_0^{M_h} \simeq_m^{M_h,M_l} S_0^{M_l}$, 
\item[(b)]
$\mathit{M_l \models \forall s. Do(\anyseqhl,S_0,s) \limp}$ \\
\hspace*{2em} $\bigwedge_{A_i \in \A^h}
    \forall \vec{x}.
 (m(\phi^{Poss}_{A_i}(\vec{x}))[s] 
    \equiv \exists s' Do(m(A_i(\vec{x})),s,s'))$,
\item[(c)] 
$M_l \models \forall s. Do(\anyseqhl,S_0,s) \limp $ \\
\hspace*{2em} $\bigwedge_{A_i \in \A^h}  \forall \vec{x}, s'. (Do(m(A_i(\vec{x})),s,s') \limp$ \\
\hspace*{4.5em} $\bigwedge_{F_i \in \F^h}  \forall \vec{y}
(m(\phi^{ssa}_{F_i,A_i}(\vec{y},\vec{x}))[s] \equiv
m(F_i(\vec{y}))[s']))$,
\end{description}
\end{small}%


\noindent
where $\phi^{Poss}_{A_i}(\vec{x})$ and $\phi^{ssa}_{F_i,A_i}(\vec{y},\vec{x})$ are as in Theorem \ref{thm:verifySound}.
\end{thm}
\textbf{Example}~Let $\D^h$ be
\[\begin{array}{l}
      Poss(A,s) \equiv True\\
      Poss(B,s) \equiv False\\
      P(do(a,s)) \equiv a = A \lor P(s)\\
      Q(do(a,s)) \equiv Q(s)\\
      R(do(a,s)) \equiv R(s)\\
      \lnot P(S_0) \land \lnot Q(S_0) \land \lnot R(S_0)
      \end{array}  \]
    and let $\D^l$ be
\[\begin{array}{l}
      Poss(A,s) \equiv True\\
      Poss(B,s) \equiv R(s)\\
      P(do(a,s)) \equiv a = A \lor P(s)\\
      Q(do(a,s)) \equiv Q(s)\\
      R(do(a,s)) \equiv (a = A \land Q(s)) \lor R(s)\\
      \lnot P(S_0) \land \lnot R(S_0)
      \end{array}  \]
  with the mapping $m$ being
  \[\begin{array}{l}
        m(A) = A, m(B) = B\\
        m(P) = P, m(Q) = Q, m(R) = R
      \end{array}  \]
Then, $\D^h$ is a complete but not sound abstraction of $\D^l$ wrt
$m$.
 $\D^h$ has a single model (up to isomorphism) $M_h$ where $Q$ is false
 initially.  We can show that $\D^l$ has a model $M_l$ that is
 $m$-bisimilar to $M_h$.  We have
 $M_l \models R(do(a,s)) \equiv R(s)$ because $Q$ remains false in all
 situations in $M_l$. Thus $R$ also remains false in all situations in
 $M_l$.  It follows that $M_l \models Poss(B,s) \equiv False$.
 $\D^h$ is not a sound abstraction of $\D^l$ wrt $m$ because $\D^l$
 also has a model $M_l'$ where $Q$ is true initially, and  $\D^h$ has
 no such model.  We have that $M_l' \not\models R(do(a,s)) \equiv
 R(s)$ and $M_l' \not\models Poss(B,s) \equiv False$.
  $\qedf$
    
  For the special case where $\D^h_{S_0}$ is a complete theory, we also have
  the following result:
  \begin{corollary} \label{cor:complete-sound-then-completeabs}
  If $\D^h_{S_0}$ is a complete theory
  (i.e., for any situation suppressed formula $\phi$,
  either $\D^h_{S_0} \models   \phi[S_0]$ or
   $\D^h_{S_0} \models   \lnot \phi[S_0]$)
   and  $\D^l$ is satisfiable,
   then if $\D^h$ is a sound abstraction of $\D^l$ with respect to $m$,
   then $\D^h$ is also a complete abstraction of $\D^l$ with respect
   to $m$.
   \end{corollary}
 



\section{Monitoring and Explanation} \label{sec:MonExp}

A refinement mapping $m$ from a high-level action theory $\D_h$ to a
low-level action theory $\D_l$ tells us what are the refinements of high-level actions into executions of low-level programs.
In some application contexts, one is interested in tracking/monitoring what
the low-level agent is doing and describing it in abstract terms,
e.g., to a client or manager.  If we have a ground low-level situation
term $S_l$ such that $\D_l \cup \{ Executable(S_l) \}$ is satisfiable,
and $\D_l \cup \{ Do(m(\vec{\alpha}),S_0,S_l) \}$ is satisfiable, then
the ground high-level action sequence $\vec{\alpha}$ is a possible way of
describing in abstract terms what has occurred in getting to situation
$S_l$.  If $\D_h \cup \{ Executable(do(\vec{\alpha},S_0)) \}$ is also
satisfiable (it must be if $\D_h$ is a sound abstraction of $\D_l$ with respect to
$m$), then one can also answer high-level queries about what may hold
in the resulting situation, i.e., whether
$\D_h \cup \{ Executable(do(\vec{\alpha},S_0)) \land
\phi(do(\vec{\alpha},S_0))\}$
is satisfiable, and what must hold in such a resulting situation, i.e.,
whether $\D_h \cup \{ Executable(do(\vec{\alpha},S_0))\} \models
\phi(do(\vec{\alpha},S_0))$.  One can also answer queries about
what high-level action $\beta$ might occur next,
 i.e., whether $\D_h \cup \{ Executable(do(\vec{\alpha}\beta,S_0))\}$
is satisfiable.

In general, there may be several distinct ground high-level
action sequences $\vec{\alpha}$ that match a ground low-level
situation term $S_l$; even if we have complete information and a
single model $M_l$ of $\D_l \cup \C$,
i.e., we may have
$M_l \models Do(m(\vec{\alpha}_1),S_0,S_l) \land
Do(m(\vec{\alpha}_2),S_0,S_l)$ and $\D_h \models \alpha_1 \neq \alpha_2$.\footnote{For example, suppose that we have two high 
level actions $A$ and $B$ with $m(A)= (C \mid D)$ and $m(B)= (D \mid E)$.
Then the low-level situation $do(D,S_0)$ is a refinement of both $A$
and $B$ (assuming all actions are always executable).}

In many contexts, this would be counter-intuitive and we would like to
be able to map a sequence of low-level actions performed by the
low-level agent back into a \emph{unique} abstract high-level action
sequence it refines, i.e., we would like to define an inverse mapping
\emph{function} $m^{-1}$.  Let's see how we can do this.
%
%
First, we introduce the abbreviation $lp_m(s,s')$,
meaning that \emph{$s'$ is a largest prefix of $s$
that can be produced by executing a sequence of high-level actions}:

\begin{small}
\[\begin{array}{l}
lp_m(s,s') \doteq 
Do(\anyseqhl,S_0,s') \land {} s' \leq s \land {}\\
\hspace*{4.5em}\forall s''. (s' < s'' \leq s \supset \lnot Do(\anyseqhl,S_0,s''))
\end{array}\]
\end{small}%
\noindent
We can show that the relation $lp_m(s,s')$ is actually 
a total function:
\begin{thm}\label{thm:unique-lp_m}
  For any refinement mapping $m$ from $\D_h$ to $\D_l$, we have that:

\begin{small}  
\begin{enumerate}
\item
$D_l \cup \C \models \forall s.\exists s'. lp_m(s,s')$,
\item
$D_l \cup \C \models \forall s \forall s_1 \forall s_2. lp_m(s,s_1)
\land lp_m(s,s_2) \supset s_1 = s_2$.
\end{enumerate}
\end{small}%
\end{thm}

\noindent
Given this result, we can introduce the notation $lp_m(s) = s'$ to
stand for $lp_m(s,s')$.


To be able to map a low-level action sequence back to a \emph{unique}
high-level action sequence that produced it, we need to assume the following
constraint:

\begin{constraint} \label{asm:firstThree}
For any distinct ground high-level action terms $\alpha$ and $\alpha'$ we have that:
 
\begin{small}
\begin{description}
\item[(a)] $\D_l \cup \C \models \forall s,s'. Do(m(\alpha), s,s') \limp$ 
$ \neg \exists \delta . Trans^*(m(\alpha'),s,\delta,s')$

\item[(b)] 
$\D_l \cup \C \models \forall s,s'. Do(m(\alpha), s,s') \limp$ 
$ \neg \exists a \exists \delta. Trans^*(m(\alpha),s,\delta,do(a,s'))$

\item[(c)] 
$\D_l \cup \C \models  \forall s,s'. Do(m(\alpha), s,s') \limp  s < s'$

\end{description}
\end{small}
\end{constraint}

\noindent
Part $(a)$ ensures that different high-level primitive actions
have disjoint sets of refinements, $(b)$ ensures that once a refinement of a 
high-level primitive action is complete, it cannot be extended further, and $(c)$ ensures
that a refinement of a high-level primitive action will produce at least one low-level action.
Together, these three conditions ensure that if we have a low-level action
sequence that can be produced by executing some high-level action
sequence, there is a unique high-level action sequence that  can produce it:
\begin{thm}\label{thm:unique-m_inv}
  Suppose that we have a refinement mapping $m$ from $\D_h$ to $\D_l$
  and that Constraint \ref{asm:firstThree} holds. Let
  $M_l$ be a model of $\D_l \cup \C$. Then for any ground situation
  terms $S_s$ and $S_e$ such that
  $M_l \models Do(\mathit{\anyseqhl},S_s, S_e)$, there exists a unique
  ground high-level action sequence $\vec{\alpha}$ such that
  $M_l \models Do(m(\vec{\alpha}),S_s,S_e)$.
\end{thm}

\noindent
Since in any model $M_l$ of $\D_l \cup \C$, for any ground situation
term $S$, there is a unique largest prefix of $S$ that can be produced
by executing a sequence of high-level actions, $S'=lp_m(S)$, and for
any such $S'$, there is a unique ground high-level action
sequence $\vec{\alpha}$ that can produce it, we can view
$\vec{\alpha}$ as the value of the inverse mapping $m^{-1}$ for $S$ in
$M_l$.
For this, let us introduce the following notation:

\begin{small}
\[\begin{array}{l}
m^{-1}_{M_l}(S) = \vec{\alpha} \doteq 
M_l \models \exists s'. lp_m(S)= s' \land
    Do(m(\vec{\alpha}),S_0,s')
\mbox{}
  \end{array}\]
\end{small}%
\noindent  
  where $m$ is a refinement mapping from $\D_h$ to $\D_l$ and
  Constraint \ref{asm:firstThree} holds, $M_l$ is a model of
  $\D_l \cup \C$, $S$ is a ground low-level situation term, and
  $\vec{\alpha}$ is a ground high-level action sequence.

Constraint \ref{asm:firstThree} however does not 
ensure that any
low-level situation $S$ can 
in fact 
be generated by executing a refinement of some
high-level action sequence; if it cannot, then the inverse mapping
will not return a complete matching high-level action sequence (e.g., we might have
$m^{-1}_{M_l}(S) = \epsilon$).
We can introduce an additional constraint that rules this
out:\footnote{One might prefer a weaker version of
Constraint \ref{asm:offMatchLL}.
For instance,  one could write a program
specifying the low level agent's possible behaviors and require that
situations reachable by executing this program can be generated by
executing a refinement of some high-level action sequence.
We discuss the use of programs to specify possible agent behaviors in the conclusion.}
 
\begin{constraint} \label{asm:offMatchLL}
\begin{small}
\[\begin{array}{l}
D_l \cup \C \models \forall s. \mathit{Executable}(s) \limp
\exists \delta.Trans^*(\mathit{\anyseqhl},S_0,\delta,s)
\end{array}\]
\end{small}%
\end{constraint}

With this additional constraint, we can show 
that for any executable low-level situation $s$, what remains after
the largest prefix that can be produced by executing a sequence of
high-level actions, i.e., the actions in the interval between $s'$ and
$s$ where $lp_m(s,s')$, can 
be generated by some (not yet
complete) refinement of a high-level primitive action:
\begin{thm}\label{thm:lp_m_rest}
If  $m$ is a refinement mapping from $\D_h$ to $\D_l$ and 
constraint \ref{asm:offMatchLL} holds, then we have that:

\begin{small}
\[\begin{array}{l}
\D_l \cup \C \models \forall s, s'. Executable(s) \land lp_m(s,s')
\supset  \exists \delta.Trans^*(\anyonehl,s',\delta,s)
\end{array}\]
\end{small}%
\end{thm}

\noindent 
\textbf{Example Cont.} Going back to the example of Section
\ref{sec:RefMap}, assume that we have complete information at the low
level and a single model $M_l$ of $\D_l^{eg}$, and suppose that the
sequence of (executable) low-level actions
$\vec{a} = [takeRoad(123,Rd_a, $ $ W,L1), takeRoad(123,Rd_b,L1, $ $L2)]$ has
occurred. The inverse mapping allows us to conclude that the
high-level action $\alpha = $ \linebreak $takeRoute(123, Rt_A, W, L2)$ has occurred,
since $m^{-1}_{M_l}(do(\vec{a},S_0)) = \alpha$.\footnote{If we do not
  have complete information at the low level, $m^{-1}_M(\vec{a})$ may
  be different for different models $M$ of $\D_l$.  To do high level
  tracking/monitoring in such cases, we need to consider all the
  possible mappings or impose additional restrictions to ensure that
  there is a unique mapping.  We leave this problem for future work.}
Since $\D_h^{eg} \models At_{HL}(123, L2, do(\alpha,S_0))$, we can also
conclude that shipment $123$ is now at location $L2$.  As well, since
$\D_h^{eg} \cup \{ Poss(takeRoute(123, Rt_B, L2, \mathit{Cf}), $ \linebreak $do(\alpha,S_0)) \}$ is
satisfiable, we can conclude that high-level action
$takeRoute($ \linebreak $123, Rt_B, L2, \mathit{Cf})$ might occur next.  Analogously, we can also
conclude that high-level action $takeRoute(123, Rt_C, L2, \mathit{Cf})$ might occur
next. However, since $\D_h^{eg} \models \neg At_{HL}(123,\mathit{Cf},do(\alpha,S_0))$, 
i.e., agent's current location is not the cafe, 
 the high-level action $Deliver(123)$ is not executable and cannot
 occur next.
$\qedf$
\section{Related Work} \label{sec:RelWork}
Given the importance of abstraction in AI, it is unsurprising that
there has been much previous work on the topic. 
%
%
%
Among logical theories of abstraction, 
Plaisted \cite{DBLP:journals/ai/Plaisted81} is perhaps the first to propose a general theory of abstraction focused on theorem proving 
(and in particular on resolution) in a first-order language.  He proposed a notion of abstraction as a mapping from a set of clauses $B$ to a 
simpler (i.e., more abstract) set of clauses $A$ that satisfied
certain properties with respect to the resolution
mechanism. Resolution proofs from $B$ map onto (possibly) simpler
resolution proofs in $A$. Plaisted introduced different types of
syntactic abstraction mappings, including \emph{renaming predicate and
  function symbols}, where several predicates (resp. functions) could
be renamed to the same predicate (resp. function) in the abstract
clause.
 Plaisted was aware that this approach could yield an abstract theory
 that produces inferences that are not sanctioned by the base theory\footnote{For example, if the base theory contains
  $JapaneseCar(x) \supset Reliable(x)$ and $EuropeanCar(x) \supset
  Fast(s)$, and we rename both $JapaneseCar$ and $EuropeanCar$ to
  $ForeignCar$, we may unjustifiably infer that a $EuropeanCar$
  instance is $Reliable$ (see
  \cite{DBLP:conf/ijcai/Tenenberg87,DBLP:conf/ijcai/NayakL95}).},
and called this issue the ``false proof'' problem; 
in fact one might even create an inconsistent abstract theory from a consistent base
theory.

%

Tenenberg \cite{DBLP:conf/ijcai/Tenenberg87} focused on
\emph{predicate mapping} abstractions, which can be considered as a
special case of Plaisted's \emph{renaming predicate and function symbols}
mapping abstractions. To ensure that the abstract theory remains
consistent, he proposed \emph{restricted predicate mappings},
where in essence, the
clauses from the base theory that distinguish the predicates being
conflated are not mapped.
%
More precisely,
this is achieved by requiring that for each clause $C$ in the abstract level, there exists a predicate (or clause) $D$ in the concrete level clause set such that $D$  maps into $C$ and either $C$ is a positive clause or for every $D$ that maps to such $C$ it is the case that D can be derived from the low-level clause set.

Giunchiglia and Walsh \cite{Giunchiglia:Walsh:1992} consider abstractions as \emph{syntactic} mappings between two representations of a problem modeled as axiomatic formal systems. A formal system is defined as a triple $\Sigma=(L,\Theta, \Delta)$, where $\Theta$ is a set of axioms in the language $L$, and $\Delta$ is the deductive machinery of $\Sigma$ (i.e., set of inference rules).  An abstraction $f: \Sigma_{base} \Rightarrow \Sigma_{abs}$ is defined as a mapping between formal systems $\Sigma_{base}$ (i.e., the \emph{ground} space) and $\Sigma_{abs}$ (i.e., the \emph{abstract} space), with languages $L_{base}$ and $L_{abs}$, respectively, and an effective, total function $f_L: L_{base} \rightarrow L_{abs}$ which is referred to as the \emph{mapping function}. 

This approach classifies abstractions based on their \emph{effects on
  provability}; that is, whether the set of theorems of the high-level
theory $\mathit{TH}(\Sigma_{abs})$ are equal to, a subset, or a superset of the
set of theorems of the low-level theory $\mathit{TH}(\Sigma_{base})$. An
abstraction $f: \Sigma_{base} \Rightarrow \Sigma_{abs}$ is said to be
$(i)$ a theorem-increasing (TI) abstraction if and only if, for any well formed formula
(wff) $\alpha$, if $\alpha \in \mathit{TH}(\Sigma_{base})$ then
$f_{L}(\alpha) \in \mathit{TH}(\Sigma_{abs})$, $(ii)$ a
theorem-decreasing (TD) abstraction if and
only if, for any wff $\alpha$, if $f_L(\alpha) \in \mathit{TH}(\Sigma_{abs})$
then $\alpha \in \mathit{TH}(\Sigma_{base})$, and $(iii)$ a
theorem-constant (TC) abstraction if it is both a TI and a TD
abstraction.
It is clear that our notions of sound abstraction, complete
abstraction, and sound and complete abstraction are related to
Giunchiglia and Walsh's notions of TD abstraction, TI abstraction, and
TC abstraction respectively, as they have similar properties with respect
to theorems/entailments.

Giunchiglia and Walsh \cite{Giunchiglia:Walsh:1992} take certain
subclasses of TI abstractions to be useful, especially
in applications where the abstract space is used to help find a proof
in the ground space.
For instance, ABSTRIPS \cite{DBLP:journals/ai/Sacerdoti74} uses a TI
abstractions that admit false proofs and are used as heuristics to guide search to speed up problem solving.
%
\cite{Giunchiglia:Walsh:1992}  reconstructed a number of previous approaches to abstraction in their framework and analyzed their properties; examples include planning with ABSTRIPS \cite{DBLP:journals/ai/Sacerdoti74}, common sense reasoning \cite{DBLP:conf/ijcai/Hobbs85}, as well as the predicate abstractions of Plaisted and Tenenberg.


Nayak and Levy \cite{DBLP:conf/ijcai/NayakL95} point out that while the
syntactic theory of abstraction proposed in
\cite{Giunchiglia:Walsh:1992} captures important aspects of different
types of abstractions, and enables their use in reasoning by theorem provers, such
a theory fails to explicitly capture the underlying justifications or assumptions that lead to the abstraction. 
Hence \cite{DBLP:conf/ijcai/NayakL95} proposed a \emph{semantic
  theory} of abstraction which focuses on a \emph{mapping between the
  models} of concrete and abstract theories. Abstraction is viewed as
a two-step process: in the first step, the intended domain model is
abstracted; and in the second step, a set of (abstract) formulas is
constructed that capture the abstracted domain model.
In general, an \emph{abstraction mapping} is a function $\pi$ that
maps a model/interpretation $M_{base}$ of a theory $T_{base}$ in the
base language $L_{base}$ to an interpretation $\pi(M_{base})$ of the
abstract language $L_{abs}$.  For first-order languages, an
abstraction mapping can be specified by giving formulas of the base
language that define the universe and the denotation of predicate and
function symbols of $\pi(M_{base})$ in terms of that in $M_{base}$:
$(i)$ a wff $\pi_{\forall}(x)$ which defines the universe of
$\pi(M_{base})$ as the set of entities that satisfy $\pi_{\forall}(x)$
in $M_{base}$, $(ii)$ a wff $\pi_R(x_1,\ldots,x_n)$ (with $n$ free
variables) for each n-ary abstract relation $R$, where the denotation of
$R$ in $\pi(M_{base})$ is the relation defined by
$\pi_R(x_1,\ldots,x_n)$ in $M_{base}$ (the set of entity $n$-tuples that
satisfy it) restricted to universe of $\pi(M_{base})$, and
$(iii)$ similar wffs that specify the denotations of 
abstract function symbols and constants.
%
%
Nayak and Levy also define the class of model increasing (MI)
abstractions, which is a strict subset of TD abstractions and yield no
false proofs.  Moreover, as an MI abstraction can be weaker than the
intended model-level abstraction, they also define the notion of
\emph{strongest} MI abstraction of a base theory.  This would yield an
abstract theory that \emph{precisely} implements the intended
model-level abstraction.
Notice that Nayak and Levy's notion of MI abstraction is closely related to our notion of
sound abstraction and their notion of strongest MI abstraction is related to
our notion of sound and complete abstraction.


Ghidini and Giunchiglia \cite{DBLP:conf/ecai/GhidiniG04} propose a semantic formalization of the notion of abstraction based on the Local Models Semantics  \cite{DBLP:journals/ai/GhidiniG01}. They associate to each of the concrete and abstract languages a set of interpretations (i.e., context). The abstraction mapping is formalized as a \emph{compatibility relation} that defines how meaning is preserved when moving from the concrete to abstract representations.

In Model theory \cite{DBLP:books/daglib/0030198,standordModelTheory},
an interpretation of a structure $A_1$ in another structure $A_2$
(whose signature may be unrelated to $A_1$) is a well established
notion that approximates the idea of representing $A_1$ inside
$A_2$. Starting with $A_2$, it is possible to build $A_1$ by defining
the domain $D_1$ of $A_1$ as well as all the labeled relations and
functions of $A_1$ as relations definable in $A_2$ by certain formulas
(with parameters). A further refinement involves finding a definable
equivalence relation on $D_1$ and considering the domain of $A_1$ to
be set of equivalence classes of this relation. A typical example is
the interpretation of the group $\mathbb{Z}$ of integers in the
structure $\mathbb{N}$ which consists of natural numbers
$0,1,2,\ldots$ with labels for $0,1$ and $+$. Interpreting structure
$A_1$ in a structure $A_2$ results in the ability to translate every
first-order statement about $A_1$ into a first-order statement about
$A_2$;
this means that 
all of $A_1$ can be read from $A_2$.
If it is possible to generalize this notion to interpreting a family
of models of a theory $T_2$, always using the same defining formulas,
then the resulting structures will all be models of a theory $T_1$
that can be read from $T_2$ and the defining formulas.
Then we can say that theory $T_1$ is interpreted in theory $T_2$.
This also shows that $T_1$ is reducible to $T_2$.


Our work is clearly distinct from all of the approaches discussed
above, which formalize abstraction of \emph{static} logical theories.
We instead focus on abstraction of \emph{dynamic} domains.
In our approach we have assumed that the high-level and the
low-level have the same domain/universe; this is mainly for
simplicity and we leave exploring object abstraction for
future work. As mentioned, Nayak and Levy
\cite{DBLP:conf/ijcai/NayakL95} handle object abstraction
by introducing a special formula that designates which of the low-level objects exist
in the high-level theory. Object abstraction is also considered in
\cite{DBLP:conf/ijcai/Hobbs85}, with the help of an
indistinguishability relation, which can be defined by means of a set
of relevant predicates (i.e., subset of predicates of the theory that
are considered to be relevant to the situation at hand). In this
approach, objects $x$ and $y$ are indistinguishable if no relevant
predicate distinguishes between them.


Abstraction techniques are widely used in model checking to deal with
very large state spaces. 
A popular approach is counterexample-guided abstraction
refinement (CEGAR)
\cite{DBLP:conf/cav/ClarkeGJLV00}, which works by
using automated techniques  to iteratively
construct and refine abstractions until a desired precision level
is reached. The approach starts by computing an abstraction which is
an upper approximation. In this setting, when a
specification
(\cite{DBLP:conf/cav/ClarkeGJLV00}  uses $ACTL^*$, a fragment of
$CTL^*$ which only permits  universal quantification over paths, as
specification language) holds in the
abstract model, it also holds in the concrete model.
However, when a
counterexample is found in the abstract model, it must be checked to
see  whether it is reproducible at the concrete level, or it is
spurious,
and needs to be excluded by refining the abstraction.
The initial abstraction is obtained by partitioning the domain of
values of variable clusters into classes that are equivalent with
respect to the satisfaction of atomic conditions that determine system
transitions or appear in the specification and using one
representative value from each equivalence class.
When a spurious counterexample is obtained, the abstraction is refined
by splitting an equivalence class to eliminate it.
The approach was extended by Shoham and Grumberg
\cite{DBLP:conf/tacas/ShohamG04} to verify arbitrary $CTL$
specifications and avoid the need to check if a potential
counterexample found in the abstract system also holds in the concrete system
by using a three-valued semantics.


The abstractions studied in this paper are quite different. They are
defined by the user rather than by some automatic technique. They
involve new high-level fluents and actions that are meaningful to the
users and can be used to express many high-level goals and
tasks/programs of interest. They can be used to speed up planning and
to give high-level explanations of system behavior.


There is also work showing decidability of model checking of temporal
properties in infinite state/object domain systems where the ``active
domain'', i.e., the set of objects that are in the extension of a
predicate in the state, remains bounded
\cite{DBLP:journals/ai/GiacomoLP16}.
Decidability is shown by constructing a finite abstract transition
system that is bisimilar to the original infinite one.
This kind of abstraction is also quite different from ours. First, it
is not defined by a designer, but it is automatically computed by
leveraging the properties of bounded theories. Second, it is
essentially based on finding a finite number of proxies for real
objects and then reusing the proxies, exploiting a locality principle
that only objects in the current active domain and in the next state's
active domain need to be distinguished.

In planning, several notions of abstraction have been investigated. 
Among the first approaches to abstraction in planning was \emph{precondition elimination abstraction}, which was initially introduced in context of ABSTRIPS \cite{DBLP:journals/ai/Sacerdoti74}. \emph{Hierarchical Task Networks} (HTNs) (e.g., \cite{DBLP:journals/amai/ErolHN96}),  abstract over a set of (non-primitive) tasks. Encodings of HTNs in \ConGolog with enhanced features like exogenous actions and online executions have been studied by Gabaldon \cite{Gabaldon:2002}. In contrast to our approach, \cite{Gabaldon:2002} uses a single BAT; also it does not provide abstraction for fluents.

Planning with \emph{macro operators} (e.g., \cite{Korf:1987}), is another approach to abstraction which represents meta-actions built from a sequence of action steps. McIlraith and Fadel \cite{McIlraith:Fadel:2002}, and Baier and McIlraith \cite{DBLP:conf/kr/BaierM06}  investigate planning with \emph{complex actions} (a form of macro actions) specified as \Golog  programs.  Differently from our approach, \cite{McIlraith:Fadel:2002,DBLP:conf/kr/BaierM06} compile the abstracted actions into a new BAT that contains both the original and abstracted actions. Also, they only deal with deterministic complex actions and do not provide abstraction for fluents. Our approach provides a refinement mapping (similar to that of Global-As-View in data integration \cite{DBLP:conf/pods/Lenzerini02}) between an abstract BAT and a concrete BAT.

Generalized planning (e.g., \cite{DBLP:conf/ijcai/Levesque05, DBLP:conf/ijcai/HuG11, DBLP:journals/ai/SrivastavaIZ11, DBLP:conf/ijcai/BonetG18}) 
studies the characterization and computation of planning solutions that
can generalize over a set of planning instances. An influential work
by Bonet and Geffner
\cite{DBLP:conf/ijcai/BonetG18} proposed a notion of \emph{abstract
  actions}, which abstract over concrete actions in different
instances of a generalized planning problem, and capture the effect of
actions on the common set of (Boolean and numerical) features.They 
show that if the abstraction is sound, then an abstract policy which
is a solution in the abstract space, can always be instantiated to
provide a solution in the original space.
%
%
Other work in the area uses abstraction, for instance,
Aguas et al. \cite{DBLP:conf/ijcai/AguasCJ16}, which proposed hierarchical
finite state controllers for generalized planning.
%
Another related work is \cite{DBLP:conf/ijcai/LuoL19}, which proposes to represent
general strategies to solve a class of possibly
infinitely many games that have similar structures
by a finite state automaton encoded in a BAT with edges labeled by \Golog
programs.  They then show how to use counterexample-guided local search for
invariants to verify that the strategy is a solution.
We discuss other generalized planning work that builds on our
abstraction framework in the next section.

Hierarchical planning approaches mainly focus on improving planning
efficiency.  Generalized planning works often use abstractions but
focus on obtaining general strategies that solve a class of planning problems.
Our work instead provides a general framework for abstracting an agent's
action repertory and how it affects the domain.  It supports various
forms of reasoning, not just planning.  It can also be used to explain
low-level behavior in high-level terms and can be applied in many ways.



\section{Adaptations and Extensions of the Abstraction Framework} \label{sec:extWork}
Since it first appeared in \cite{DBLP:conf/aaai/BanihashemiGL17}, our
abstraction framework has been adapted, used, and extended in various
directions.  We summarize such work in this section.

Some work has addressed the issue of  automated verification and synthesis of abstractions.
In \cite{DBLP:conf/ecai/Luo0LL20}, Luo et al.\ investigate the relationship between
our account of agent abstraction and the well-known notion of
\emph{forgetting} in first-order/second-order logic.
They also address the problem of \emph{synthesizing a sound and complete abstraction given
a low-level BAT and a refinement mapping}.
They show that one can use forgetting (of low-level fluent and action
symbols) to obtain second-order theory that is a sound and complete
abstraction of a low-level BAT
for a given mapping, provided that the mapping is \emph{nondeterministic
uniform (NDU)} with respect to the low-level BAT, i.e, such that different refinements
of a high-level action sequence are indistinguishable in the
high-level language.\footnote{Note that it follows our Theorem \ref{thm:verifySoundComplete} that their NDU
  condition is satisfied whenever a sound abstraction BAT exists for the given low-level BAT and mapping.}
%
%
In general, the result of this method may not have the form of a BAT.
%
But it is also shown that if the mapping is ``Markovian'' with respect to the low-level
BAT, i.e., such that the executability and effects of refinements of
high-level actions only depends the values of the high-level fluents
in the current situation, whenever such a situation is the result of
executing a refinement of high-level action sequence, then a sound and
complete abstraction in the form of a generalized BAT, i.e., where the
initial database, action preconditions and successor state
descriptions can be second-order formulas, can be obtained.
Finally, they show that in the propositional case, under the Markovian
restriction, a sound and complete abstraction is always computable.

Luo \cite{DBLP:conf/aaai/Luo23} studies automated verification of the
existence of a propositional agent abstraction given a low-level
finite-state labeled transition system (LTS) and a refinement mapping
that maps high-level atoms into low-level state formulas and
high-level actions into loop-free low-level programs.  Notions of
sound/complete abstractions over such LTS are defined.  Note that
actions in the high-level LTS need not be deterministic as in our
approach.  Luo then shows that the propositional agent abstraction
existence problem (for both deterministic and non-deterministic LTSs)
can be reduced to a CTLK (an extension of CTL with epistemic
operators \cite{DBLP:conf/atal/PenczekL03}) model checking problem.
The idea behind this is that low-level states reached by executions of the same
high-level actions are epistemically indistinguishable for the high-level agent.
Given the low-level LTS and mapping, an induced epistemic transition
system (ETS) is obtained.  It is shown that a propositional agent
abstraction exists if this ETS satisfies certain properties.  If the
low-level LTS is represented symbolically as a propositional STRIPS
planning domain, a symbolic induced ETS can also be obtained.  However this
symbolic ETS includes all the propositional variables that were
present in the given low-level LTS, and so is it not fully abstract.  The
approach has been implemented in a system that uses the
MCMAS \cite{DBLP:journals/sttt/LomuscioQR17} model checker.
Experiments on several domains from classical planning show that the
system can verify the existence of sound and complete propositional
abstractions in reasonable time for medium size domain instances.
Synthesizing abstract LTS/planning domains is left for future work.

As mentioned in the previous section, much work in \emph{generalized
  planning} exploits abstraction to generate a (typically iterative)
general plan/strategy that solves a 
set of similar planning problems.
Inspired by \cite{DBLP:conf/ijcai/BonetG18},
Cui et al.~\cite{DBLP:conf/ijcai/Cui0L21} extend our framework to
develop a uniform abstraction framework for generalized planning.
They consider abstractions that involve nondeterministic actions as in
fully-observable nondeterministic (FOND) planning, and seek strong cyclic
solutions that achieve the goal under a fairness assumption.
They use \Golog programs to represent non-deterministic actions.
To deal with the issue of termination of generalized plans under
fairness constraints, they use an extended
situation calculus with infinite histories \cite{DBLP:journals/amai/SchulteD04}.
The framework is
also extended with counting (following \cite{DBLP:conf/lics/KuskeS17})
and transitive closure (following
\cite{DBLP:conf/cav/ImmermanV97}).
In this setting, a generalized planning problem (GPP) is formalized as a triple of a BAT, a trajectory constraint and a goal.
They define notions of sound/complete abstractions with respect
to a mapping $m$ at the
level of models by introducing notions of $m$-simulation and $m$-back
simulation that are weaker than $m$-bisimulation.\footnote{Such
  notions are interesting and allow one to get interesting results
  about the existence of plans in models, but $m$-bisimulation is
  a simpler and more intuitive notion.  If one has perfect information and a
single model, then one would expect to have $m$-bisimilar models at
different levels of abstraction. If instead one has imperfect
information and multiple models, but the information one has at
the high and low levels is consistent, then our notion of
sound (and possibly also complete) abstraction defined in terms of
$m$-bisimulation seems sufficient to get the results that we want.
More analysis of these issues would be worthwhile.}
They then use these to define notions of sound/complete
abstractions (at the theory level) for GPPs.
They show that if a high-level
GPP is a  sound abstraction of a low-level GPP,
and a high-level program $\delta$ is a strong solution to the
high-level GPP, then $m(\delta)$ is a strong solution to the low-level
GPP, where $m(\delta)$ is obtained by replacing all high-level fluents and actions
in $\delta$ by their mapped value. To be a strong solution to a GPP, a program $\delta$ must be
guaranteed to achieve the goal under the trajectory constraint unless
the execution aborts/blocks.  Note that this is a rather weak notion
of strong solution given
that $\delta$ and $m(\delta)$ may be nondeterministic and have
executions that block..
A number
of existing approaches to generalized planning such as
\cite{DBLP:conf/ijcai/BonetG18} are formalized in the proposed
abstraction framework and compared.

In later work, Cui et al.~\cite{DBLP:journals/corr/abs-2205-11898}
extended this approach to support automatic verification that one has
a sound abstraction for a GPP, where the abstraction model is a 
qualitative numerical planning (QNP) problem, a common approach.
They assume that the QNP abstraction is ``bounded'', i.e., integer variables can only be incremented or decremented by one.
To do this, they first obtain a ``proof-theoretic'' characterization of their
notion of sound abstraction for GPPs (along the lines of our
Theorem \ref{thm:verifySound}),
and then identify
a sufficient condition for having a sound abstraction when
high-level action implementations are deterministic and loop-free.
They then develop a sound abstraction verification system that checks
the sufficient condition based on the SMT solver $Z3$.
Experiments showed that the system was able to successfully
verify hand-crafted QNP abstractions for several GPP domains in reasonable time.

In this paper, we have ignored the issue of sensing and knowledge acquisition.
Banihashemi et al.~\cite{DBLP:conf/ijcai/BanihashemiGL18} extend our
abstraction framework to the case where the agent is executing
\emph{online}
i.e., may acquire new knowledge while executing (e.g., by sensing)
\cite{DeGiacomoLevesqueLFCA99,DBLP:journals/amai/SardinaGLL04}. This means that the knowledge base that the agent uses in its reasoning needs to be updated during the execution. 
%
A sufficient property is identified which allows \emph{sound abstraction to persist in online executions}. This property ensures that the low level has learned as much as the high level did when a refinement of the high-level action was performed, which ensures that every low-level model still has an m-bisimilar high-level model.
%
Abstraction can also be exploited to support planning for agents that execute online. To achieve its goals, such an agent may need to
select different courses of action depending on the results of sensing actions (or the occurrence of exogenous actions).  
This work adapts definition of ability to perform a task/achieve a goal (e.g., \cite{Moore:1985,DBLP:journals/sLogica/LesperanceLLS00}) to its model of online executions. To be \emph{able} to achieve a goal, an agent needs to have a \emph{strategy} that ensures reaching the goal no matter how the environment behaves and how sensing turns out.
Ability is similar to the concept of \emph{conditional} or \emph{contingent planning} \cite{DBLP:journals/ai/CimattiPRT03,DBLP:conf/flairs/AlfordKNRG09,DBLP:series/synthesis/2013Geffner}. The main result of this work shows that under some reasonable assumptions, if one has a sound abstraction and the agent \emph{has a strategy by which it is able to achieve a goal at the high level, then one can refine it into a low-level strategy} by which the agent is able to achieve the refinement of the goal.  Furthermore, the low-level strategy can be obtained piecewise/incrementally, by finding a refinement of each step of the high-level strategy individually.  This makes reasoning about agents' abilities much easier.

Many agents operate in nondeterministic environments where the agent does not fully control the outcome of its actions (e.g., flipping a coin where the outcome may be heads or tails). In more recent work, \cite{DBLP:conf/ijcai/BanihashemiGL23,DBLP:journals/corr/abs-2305-14222} extend our framework to abstract the behavior of an agent that has nondeterministic actions based on the nondeterministic situation calculus \cite{DBLP:conf/kr/GiacomoL21} action theories.  In this setting, each high-level action is considered as being
composed of an agent action and an (implicit) environment reaction. As a consequence, the agent action (without
the environment reaction) is mapped into a low-level agent program that appropriately reflects the nondeterminism of the environment,
and the complete high-level action, including both the agent action and the environment reaction, is mapped into a low-level
system program that relates the high-level environment reaction to the low-level ones. A constraint is defined that ensures mapping is proper, i.e., agent actions and system actions are mapped in a consistent way. This allows our notion of $m$-bisimulation to extend naturally to this new setting.
This new setting supports strategic reasoning and
strategy synthesis, by allowing one to quantify separately on agent actions and environment reactions. It is shown that if the agent has a (strong FOND) plan/strategy to achieve a goal/complete a task at the abstract level, and it can always execute the nondeterministic abstract actions to completion at the concrete level, then there exists a refinement of it that is a (strong FOND) plan/strategy to achieve
the refinement of the goal/task at the concrete level.

Our framework was also recently extended along similar lines to
abstract agent behavior in multi-agent synchronous games and support
verification of strategic properties in \cite{LesperanceGR0-AAAI24}.

Related to this is work on using BAT abstraction in supervisory control of agents.
In 
\cite{DBLP:conf/atal/BanihashemiGL18}, constraints on the
agent's behavior are specified in the language of the abstract BAT. A
high-level maximally permissive supervisor (MPS) is first obtained that
enforces these constraints on the behavior of a high-level agent while
leaving the agent with as much freedom to act/autonomy as possible. The task
is then to synthesize an MPS for the concrete agent. To support
mapping an abstract supervision specification to a concrete one,
\cite{DBLP:conf/atal/BanihashemiGL18} extend the refinement mapping
to map any situation-determined high-level program
to a situation-determined low-level program that implements it
(concurrency is allowed at the high level, but the implementations of
high-level actions cannot be interleaved to maintain bisimulation).  A
low-level MPS based on the concrete specification is then obtained,
and it is shown that the high-level MPS is the abstract version of the
low-level MPS. Moreover, a hierarchically synthesized MPS may be
obtained by taking the abstract MPS and refining its actions piecewise.

Belle \cite{DBLP:journals/kbs/Belle20} extends our abstraction approach to probabilistic relational models, assuming a weighted model counting reasoning infrastructure, and focusing on static models where there are no actions.  He assumes that one defines a refinement mapping that associates each high-level atom to a (propositional) formula over the language of the low-level model/theory.  Isomorphism between a high-level model and a low-level model relative to a mapping is defined essentially as in our account.
This is then used to define a notion of weighted sound abstraction
relative to a mapping between such models which ensures that
high-level sentences that have a non-zero probability at the low level must
also have a non-zero probability at the high level, and high-level
sentences that have probability 1 at the high level must also have
probability 1 at the low level.  A weighted complete abstraction ensures the
converse.  Finally, a weighted exact abstraction is a weighted sound
and complete abstraction where the probability of any high-level
sentence is the same at the high level and low level.  The paper also discusses techniques for automated synthesis of weighted abstractions.

Hofmann and Belle \cite{DBLP:conf/atal/HofmannB23} extend this approach to support abstraction of probabilistic dynamical systems.  The high-level and low-level action theories are defined in the DSG logic (which extends DS \cite{DBLP:conf/ijcai/BelleL17}), a first-order modal version of the situation calculus that supports probabilistic belief ($B(\alpha:r)$ means that $\alpha$ is believed with probability $r$), stochastic/noisy observations and actions, and \Golog programs ($[\delta] \alpha$ means that $\alpha$ holds after every execution of program $\delta$).  Their notion of $m$-bisimulation relates epistemic states in the high-level and low-level theories; the base condition is epistemic $m$-isomorphism,  which requires not only that the states be $m$-isomorphic, but also that the degrees of belief in high-level sentences be exactly the same.  It is then shown that $m$-bisimilar states satisfy the same bounded (i.e., without the "always" operator) high-level DSG formulas. This is then extended to theories by defining  notions of sound/complete abstraction as usual.  The framework is used to show one can abstract away the stochastic aspects of the domain and obtain a non-probabilistic high-level domain theory while keeping guarantees that refinements of the high-level actions and plans achieve their goals.  One limitation is is that DSG does not support strategic reasoning: one cannot say that the agent has a strategy to execute a program to termination and achieve a goal no matter how the environment behaves, i.e., no matter how it selects sensing results and action outcomes.  The requirement that degrees of belief in high-level sentences be exactly the same in epistemic $m$-isomorphism may also be too strict in some applications.


Applying abstraction to smart manufacturing, De Giacomo et al. \cite{DBLP:journals/ai/GiacomoFLPS22} focus on manufacturing as a service in which manufacturing facilities ``bid'' to produce (previously unseen and complex) products. A facility, consists of a set of configurable manufacturing resources (e.g., robots, tools, etc.). Given a product recipe, which outlines the abstract (i.e., resource independent) tasks required to manufacture a product, a facility decides whether to bid for the product if it can synthesize, ``on the fly'', a process plan controller that delegates abstract manufacturing tasks in the supplied process recipe to the appropriate manufacturing resources.
The operations in manufacturing processes are described by basic action theories, and the processes as \ConGolog programs over such action theories. The facility basic action theory is obtained by combining the resources' basic action theories, and the facility process results from the concurrent, synchronous execution of the processes of each resource. Moreover, another basic action theory represents a common, resource independent information model of the data and objects that the recipes can manipulate and based on which they are designed. A set of mappings relate the common abstract resource-independent BAT with the resource-dependent facility BAT. Based on this representation, this work formalizes when a process recipe can be realized by facility. Two decidable cases of finite domains and bounded action theories are identified, for which techniques to actually synthesize the controller are provided.






\section{Conclusion} \label{sec:Conclusion}
In this paper, we proposed a general framework for agent abstraction
based on the situation calculus and \ConGolog.
We defined a notion of a high-level basic action theory (BAT) being a
\emph{sound/complete abstraction} of a low-level BAT with
respect to a given mapping between their respective languages.
This formalization involved the existence
of a \emph{bisimulation relation relative to the mapping} between models of
the abstract and concrete theories. Furthermore, we identified a set
of necessary and sufficient conditions for checking if one has a sound
and/or complete abstraction with respect to a mapping.
We have shown that sound abstractions have many useful
properties that ensure that we can reason about the agent's actions
(e.g., executability, projection, and planning) at the abstract level,
and refine and concretely execute them at the low level.
Finally, we identified a set of BAT constraints that ensure that for any low-level action sequence, there is a unique high-level action sequence that it refines. This is 
useful for monitoring and providing high-level explanations of behavior. 
Our framework is based on the situation calculus, which provides a
first-order representation of the state and and can model the
data/objects that the agent interacts with.


In this work, for simplicity, we focused on a single layer of abstraction,
but the framework supports extending the hierarchy to more levels.
%
Our approach can also support the use of \ConGolog programs to specify
the possible behaviors of the agent at both the high and low level, as
we can follow \cite{DBLP:conf/aaai/GiacomoLPS16} and ``compile'' such a
program into the BAT $\D$ to get a new BAT $\D'$ whose executable
situations are exactly those that can be reached by executing the
program.  





In future work, we plan to investigate how one can construct
abstraction mappings and abstract action theories. This could be done
in the context of designing a new agent system or to enhance an
existing one, for instance to support monitoring, adaptation, and
evolvability.  Note that there may be many different abstractions of a
concrete action theory, each of which may be useful for a different
purpose.  We need to identify how such requirements would be
specified.  Once we have such requirements, one could generate an
abstract language/ontology, i.e., fluents and action types, and a
mapping for them that satisfies the requirements.
After that, one could check that there exists a high-level theory that is a sound
and/or complete abstraction for such a mapping and a given low-level
theory, and if so, synthesize one and verify that it is indeed a
sound and/or complete abstraction.  Many of these steps could be
automated to a significant extent.  But a human designer would likely
need to be involved to refine/revise the requirements and help adjust
the high-level language and mapping to ensure that a sound/complete
abstraction is obtained.  For example, if different refinements of a
high-level action can produce situations that are not $m$-isomorphic,
then we can either switch to more abstract high-level fluents to
ensure the resulting situations are indeed $m$-isomorphic, or split the
high-level action into several for which we do have $m$-isomorphic
results.  We plan to investigate methodologies and develop tools for
this process.


As mentioned in the previous section, there has been some work on checking
whether a sound/complete abstraction exists for a given low-level BAT
and mapping, and on synthesizing a high-level BAT when it does,
focusing on the propositional case
\cite{DBLP:conf/ecai/Luo0LL20,DBLP:conf/aaai/Luo23}.
For verifying that a high-level BAT is a sound abstraction of a
low-level BAT with respect to a mapping in the general infinite-states
case, one can try to use general first-order and higher-order logic
theorem proving techniques and tools; for instance, one could build
upon Shapiro's work \cite{DBLP:conf/atal/ShapiroLL02}. Another
possibility is to restrict attention to bounded basic action theories
\cite{DBLP:journals/ai/GiacomoLP16}. These are action theories where
it is entailed that in all situations, the number of object tuples
that belong to the extension of any fluent is bounded, although the
object domain remains infinite and an infinite run may involve an
infinite number of objects.  In \cite{DBLP:journals/ai/GiacomoLP16},
it is shown that verifying a large class of temporal properties over
bounded BATs is decidable.
Pre-trained large language models have also been used to help in defining
planning domains
\cite{DBLP:journals/corr/abs-2305-14909,DBLP:journals/corr/abs-2404-07751}.
Perhaps one could use them to help generate useful abstractions.



As already mentioned, we have already extended our
agent abstraction framework to deal with nondeterministic domains and
contingent planning
\cite{DBLP:conf/ijcai/BanihashemiGL23,DBLP:journals/corr/abs-2305-14222},
as well as agents that perform sensing and acquire new knowledge at execution
time \cite{DBLP:conf/ijcai/BanihashemiGL18}.  Further work in these
areas is indicated, in particular to support multiple environment
models involving a range of contingencies, as well as strategic
reasoning in multi-agent environments.
%
%
We would also like to explore how agent abstraction can be used in verification,
where there is some related work \cite{DBLP:conf/ecai/MoL016,DBLP:conf/ecai/BelardinelliLM16}.







\section*{Acknowledgements}
This work has been partially supported by the ERC Advanced Grant WhiteMech (No. 834228),
the National Science and Engineering Research Council of Canada, and York University.












\bibliographystyle{elsarticle-harv}
\bibliography{citations}

\appendix

\section{Proofs}

\subsection{$m$-Bisimulation}

\textbf{Lemma} \ref{lem:sitSupBisim} 
If $s_h \simeq_m^{M_h,M_l} s_l$, then for any high-level
  situation-suppressed formula $\phi$, we have that:
\[\begin{array}{l}
M_h,v[s/s_h] \models \phi[s]\ \ \mbox{if and only if}\ \ 
M_l,v[s/s_l] \models m(\phi)[s]
\end{array}\]

\textbf{Proof:} By induction of the structure of $\phi$, using the definition of $m$-isomorphic situations. \qed

\textbf{Lemma} \ref{lem:bisimL2HOffND}
If  $M_h \sim_m M_l$, then
for any sequence of high-level actions $\vec{\alpha}$,
we have that

\begin{small}  
\[\begin{array}{l}
\mbox{if } M_l,v[s'/s_l] \models Do(m(\vec{\alpha}),S_0,s'),
    \mbox{ then there exists $s_h$ such that:}\\
    M_h,v[s'/s_h]  \models s_h = do(\vec{\alpha},S_0) \land
    Executable(s_h) \mbox{ and } s_h \sim_m^{M_h,M_l} s_l\\[1ex]
    \mbox{and } \\[1ex]
\mbox{if }  M_h,v[s'/s_h]  \models s_h = do(\vec{\alpha},S_0) \land
    Executable(s_h)\\
\mbox{then there exists $s_l$ such that } M_l,v[s'/s_l] \models Do(m(\vec{\alpha}),S_0,s')
 \mbox{ and } s_h \sim_m^{M_h,M_l} s_l.
\end{array}\]
\end{small}%

\noindent
\paragraph{Proof} The result follows easily by induction on the
length of $\alpha$ using the definition of $m$-bisimulation. \qed


\noindent
\textbf{Theorem} \ref{thm:bisimL2HOff} 
  If  $M_h \sim_m M_l$, then 
  for any sequence of ground high-level actions $\vec{\alpha}$ and
  any high-level situation-suppressed formula $\phi$,
we have that:

  \begin{small}  
  \[\begin{array}{l}
  M_l \models \exists s' Do(m(\vec{\alpha}),S_0,s') \land
  m(\phi)[s'] \hspace{1em} \mbox{ if and only if }\\
  \hspace{4em} M_h \models Executable(do(\vec{\alpha},S_0)) \land
  \phi[do(\vec{\alpha},S_0)].
   \end{array}\]
   \end{small}%



\noindent
\paragraph{Proof} The result follows immediately from Lemma
\ref{lem:sitSupBisim}  and Lemma
\ref{lem:bisimL2HOffND}.  \qed

\subsection{Sound Abstraction}

\textbf{Theorem} \ref{Thm:ExecutableL2HOffEntail}
Suppose that $\D_h$ is a sound abstraction of $\D_l$ relative to 
mapping $m$.  Then for any ground high level action sequence $\vec{\alpha}$ and
for any high level situation suppressed formula $\phi$,
if $\D_h \models Executable(do(\vec{\alpha},S_0)) \land \phi[do(\vec{\alpha},S_0)]$, then
$\D_l \cup \C \models \exists s. Do(m(\vec{\alpha}),S_0,s) \land m(\phi)[s]$.

\textbf{Proof:} 
Assume that $\D_h$ is a sound
abstraction of $\D_l$ wrt $m$ and that  $D_h \models
Executable(do(\vec{\alpha},S_0)) \land \phi[do(\vec{\alpha},S_0)]$.
Take an arbitrary model $M_l$ of $\D_l \cup \C $.  Since $\D_h$ is a sound
abstraction of $\D_l$ wrt $m$, there exists a model $M_h$ of $\D_h$
such that $M_h \sim_m M_l$.  Given our assumption,  it follows that
$M_h \models
Executable(do(\vec{\alpha},S_0)) \land \phi[do(\vec{\alpha},S_0)]$.
It then follows by Theorem \ref{thm:bisimL2HOff} that
$M_l \models \exists s. Do(m(\vec{\alpha}),S_0,s) \land m(\phi)[s]$.
$M_l$ was an arbitrarily chosen model of $\D_l \cup \C$, so the result follows.
\qed

\noindent
\textbf{Corollary} \ref{cor:HLdeterminismOfLL}
If  $\D_h$ is a sound abstraction of $\D_l$ relative to
mapping $m$, then for any sequence of ground high-level actions
$\vec{\alpha}$ and for any high-level situation-suppressed formula
$\phi$, we have that:
\[\begin{array}{l}
\D_l \cup \C \models Do(m(\vec{\alpha}),S_0,s) \land
Do(m(\vec{\alpha}),S_0,s') \limp
    (m(\phi)[s] \equiv m(\phi)[s'])
\end{array}\]

\textbf{Proof:} By contradiction.  
Suppose that there exist $M_l$ and $v$ such that 
$M_l,v \models \D_l \cup \C \cup \{ Do(m(\vec{\alpha}),S_0,s) \land
Do(m(\vec{\alpha}),S_0,s') \land m(\phi)[s] \land \neg m(\phi)[s']\}$.
Since $\D_h$ is a sound abstraction of $\D_l$ relative to
mapping $m$, by Corollary \ref{cor:ExecutableL2HOffSat}, it follows that
$\D_h \cup
\{ Executable(do(\vec{\alpha},S_0)) \land \phi[do(\vec{\alpha},S_0)]
\land \neg \phi[do(\vec{\alpha},S_0)] \}$ is satisfiable, a
contradiction.
\qed

\noindent
\textbf{Theorem} \ref{thm:ExecutableL2HOffEntail2}
If  $\D_h$ is a sound abstraction of $\D_l$ relative to
mapping $m$, then for any sequence of ground high-level actions
$\vec{\alpha}$ and for any ground high-level action $\beta$, we have that:
\[\begin{array}{l}
\D_l \cup \C \models 
\exists s. Do(m(\vec{\alpha}\beta),S_0,s) \supset
    (\forall s. Do(m(\vec{\alpha}),S_0,s) \limp \exists s'. Do(m(\beta),s,s'))
\end{array}\]

\textbf{Proof:} Take an arbitrary model $M_l$ of $\D_l \cup \C$ and
valuation $v$ and assume that
$M_l,v \models \exists s. Do(m(\vec{\alpha}\beta),S_0,s)$.  It follows
that there exists $s_l$ such that
$M_l,v[s/s_l ]\models Do(m(\vec{\alpha}),S_0,s) \land \exists
s'. Do(m(\beta),s,s')$.
Since $\D_h$ is a sound abstraction of $\D_l$ wrt $m$, there exists a
model $M_h$ of $\D_h$ such that $M_h \sim_m M_l$.
Then by Lemma \ref{lem:bisimL2HOffND}, it follows that
there exists a situation $s_h$ such that 
$M_h, v[s/s_h]  \models s = do(\vec{\alpha},S_0) \land Executable(s) \land
Poss(\beta,s)$.
Take an arbitrary situation $s'_l$ and suppose that
$M_l,v[s/s'_l] \models Do(m(\vec{\alpha}),S_0,s)$.  Then it follows by
Lemma \ref{lem:bisimL2HOffND} 
that $s_h \sim_m^{M_h,M_l} s'_l$.
Since we also have that 
$M_h \models Poss(\beta,do(\vec{\alpha},S_0))$, it follows that
$M_l,v[s/s'_l] \models \exists s'. Do(m(\beta),s,s')$.  Since $s'_l$
was chosen arbitrarily, it follows that
$M_l,v \models \forall s. Do(m(\vec{\alpha}),S_0,s) \limp \exists
s'. Do(m(\beta),s,s')$.  \qed

To prove Theorem \ref{thm:verifySound} (and Theorem \ref{thm:verifyComplete}), we start by
showing some lemmas.

\begin{lem} \label{lem:verifySound1}
If $M_h \models \D^h$ for some high-level BAT  $\D^h$ and $M_l
\models \D^l \cup \C$ for some low-level BAT  $\D^l$ and $M_h
\sim_m M_l$ for some mapping $m$, then 
\begin{small}
\begin{description}
\item[(a)]
$\mathit{M_l \models \forall s. Do(\anyseqhl,S_0,s) \limp}$ \\
\hspace*{2em} $\bigwedge_{A_i \in \A^h}
    \forall \vec{x}.
 (m(\phi^{Poss}_{A_i}(\vec{x}))[s] 
    \equiv \exists s' Do(m(A_i(\vec{x})),s,s')),$
\item[(b)]
$M_l \models \forall s. Do(\anyseqhl,S_0,s) \limp $ \\
\hspace*{2em} $\bigwedge_{A_i \in \A^h}  \forall \vec{x}, s'. (Do(m(A_i(\vec{x})),s,s') \limp$ \\
\hspace*{4.1em} $\bigwedge_{F_i \in \F^h}  \forall \vec{y}  (m(\phi^{ssa}_{F_i,A_i}(\vec{y},\vec{x}))[s] \equiv m(F_i(\vec{y}))[s'])),$
\end{description}
\end{small}%
\noindent
where $\phi^{Poss}_{A_i}(\vec{x})$ is the right hand side of the precondition axiom for action $A_i(\vec{x})$, and $\phi^{ssa}_{F_i,A_i}(\vec{y},\vec{x})$ is the right hand side of the successor state axiom for $F_i$ instantiated with action
$A_i(\vec{x})$ where action terms have been eliminated using $D^h_{ca}$. 
\end{lem}

\textbf{Proof}
By contradiction.
Assume that $M_h$ is a model of a high-level BAT  $\D^h$ and $M_l$ is a
model of a low-level BAT  $\D^l \cup \C$ and $M_h \sim_m M_l$.
Suppose that condition (a) does not hold.
Then there exists a ground high-level
action sequence $\vec{\alpha}$, a ground low-level situation term $S$,
and a ground high-level action $A_i(\vec{x})$ such that
$M_l \models Do(m(\vec{\alpha}), S_0, S)$ and 
either (*)
$M_l \models m(\phi^{Poss}_{A_i}(\vec{x}))[S]$ and $M_l \not\models \exists s'. Do(m(A_i(\vec{x})),S,s')$
or (**)
$M_l \not\models m(\phi^{Poss}_{A_i}(\vec{x}))[S]$ and $M_l \models
\exists s'. Do(m(A_i(\vec{x})),S,s')$.
In case (*), by Theorem \ref{thm:bisimL2HOff}, since $M_h \sim_m M_l$,
 it follows that $M_h \models Executable(do(\vec{\alpha}, S_0)) \land
\phi^{Poss}_{A_i}(\vec{x})[do(\vec{\alpha}, S_0)]$. 
Since $M_h \models \D^h_{Poss}$, we must also have that 
$M_h \models Poss(A_i(\vec{x}),do(\vec{\alpha}, S_0))$, and thus that
$M_h \models
Executable(do([\vec{\alpha},A_i(\vec{x})], S_0))$.
Thus by Theorem \ref{thm:bisimL2HOff}, $M_l \models Do(m(\vec{\alpha}), S_0, S) \land
\exists s' Do(m(A_i(\vec{x})),S,s')$, which contradicts (*).
Case (**) can be shown to to lead to a contradiction by a similar
argument.
\\[1ex]
Now suppose that condition (b) does not hold.
Then there exists a ground high level
action sequence $\vec{\alpha}$, a ground high-level action
$A_i(\vec{x})$, and ground low-level situation terms $S$ and
$S'$ such that
$M_l \models Do(m(\vec{\alpha}), S_0, S) \land $ \linebreak $ Do(m(A_i(\vec{x})), S, S')$ and 
either (*)
$M_l \models m(\phi^{ssa}_{F_i,A_i}(\vec{y},\vec{x}))[S]$ and \linebreak $M_l \not\models m(F_i(\vec{y}))[S']$
or (**)
$M_l \not\models m(\phi^{ssa}_{F_i,A_i}(\vec{y},\vec{x}))[S]$ and $M_l \models
m(F_i(\vec{y}))[S']$.
In case (*), by Theorem \ref{thm:bisimL2HOff}, since $M_h \sim_m M_l$,
it follows that \linebreak $M_h \models Executable(do(\vec{\alpha}, S_0)) \land
\phi^{ssa}_{F_i,A_i}(\vec{y},\vec{x})[do(\vec{\alpha}, S_0)] \land
Poss(A_i(\vec{x}),do(\vec{\alpha}, S_0))$.
Since $M_h \models \D^h_{ssa}$, we must also have that 
$M_h \models F_i(\vec{y})[do([\vec{\alpha},A_i(\vec{x})], S_0)]$.
Thus by Theorem \ref{thm:bisimL2HOff}, $M_l \models Do(m(\vec{\alpha}), S_0, S) \land
Do(m(A_i(\vec{x})),S,S') \land $ \linebreak $ m(F_i(\vec{y}))[S']$, which contradicts (*).
Case (**) can be shown to to lead to a contradiction by a similar
argument.
\qed

\noindent
The above lemma implies that if $\D^h$ is a sound abstraction of $\D^l$ wrt $m$, 
then $\D^l$ must entail the mapped high-level successor state axioms
and entail that the mapped conditions for a high-level action to be executable
(from the precondition axioms of $\D^h$) correctly capture the
executability conditions of their refinements.

We also prove another lemma:
\begin{lem} \label{lem:verifyComplete1}
Suppose that $M_h \models \D^h$ for some high-level BAT  $\D^h$ and $M_l
\models \D^l \cup \C$ for some low-level BAT  $\D^l$ and $m$ is a
mapping between the two theories.
Then if\\[1ex]
\textbf{(a)}~~$S_0^{M_h} \simeq_m^{M_h,M_l} S_0^{M_l}$,\\[0.8ex]
\textbf{(b)}~~$\mathit{M_l \models \forall s. Do(\anyseqhl,S_0,s) \limp}$ \\
\hspace*{4.5em} $\bigwedge_{A_i \in \A^h}
    \forall \vec{x}.
 (m(\phi^{Poss}_{A_i}(\vec{x}))[s] 
    \equiv \exists s' Do(m(A_i(\vec{x})),s,s'))$ and \\[0.8ex]
\textbf{(c)}~~$M_l \models \forall s. Do(\anyseqhl,S_0,s) \limp $ \\
\hspace*{4.5em} $\bigwedge_{A_i \in \A^h}  \forall \vec{x},
s'. (Do(m(A_i(\vec{x})),s,s') \limp$ \\
\hspace*{7em} $\bigwedge_{F_i \in \F^h}  \forall \vec{y}
(m(\phi^{ssa}_{F_i,A_i}(\vec{y},\vec{x}))[s] \equiv
m(F_i(\vec{y}))[s'])),$\\[0.8ex]
then $M_h \sim_m M_l$,\\[1ex]
where $\phi^{Poss}_{A_i}(\vec{x})$ is the right hand side of the precondition axiom for action $A_i(\vec{x})$, and $\phi^{ssa}_{F_i,A_i}(\vec{y},\vec{x})$ is the right hand side of the successor state axiom for $F_i$ instantiated with action
$A_i(\vec{x})$ where action terms have been eliminated using $D^h_{ca}$. 
\end{lem}

\textbf{Proof:}
Assume that the antecedent holds.
Let us show that $M_h \sim_m M_l$.
Let $B$ be the relation over $\Delta_S^{M_h} \times \Delta_S^{M_l}$
such that
\begin{center}
\begin{tabular}{l}
$\langle s_h, s_l \rangle \in B$\\
\qquad if and only if\\
there exists
a ground high-level action sequence $\vec{\alpha}$\\ such that
 $M_l,v[s/s_l] \models Do(m(\vec{\alpha}),S_0,s)$\\
 and $s_h = do(\vec{\alpha},S_0)^{M_h}$. 
\end{tabular}
\end{center}
Let us show that $B$ is an $m$-bisimulation relation between $M_h$ and $M_l$.  
%
We need to show that if $\langle s_h, s_l \rangle \in B$, then it satisfies
the three conditions in the definition of $m$-bisimulation.
We prove this by induction on $n$, the number of actions in $s_h$.
\\
Base case $n=0$:
By the definition of $B$, $s_h = S_0^{M_h}$, $\alpha = \epsilon$, and $s_l = S_0^{M_l}$.
By (a) we have that
$S_0^{M_h} \simeq_m^{M_h,M_l} S_0^{M_l}$,
so condition 1 holds.
By Lemma \ref{lem:sitSupBisim}, it follows that $M_h,v[s/s_h] \models
\phi^{Poss}_A(\vec{x})[s]$ if and only if $M_l,v[s/s_l] \models
m(\phi^{Poss}_A(\vec{x}))[s]$ for any high-level primitive action type
$\mathit{A} \in \mathit{\A_h}$.  Thus by the action precondition axiom
for $A$, $M_h,v[s/s_h] \models
Poss(A(\vec{x}),s)$ if and only if $M_l,v[s/s_l] \models
m(\phi^{Poss}_A(\vec{x}))[s]$.  By condition (b), we have that 
$M_l,v[s/s_l] \models m(\phi^{Poss}_A(\vec{x}))[s]$ if and only if
$M_l,v[s/s_l] \models \exists s'. Do(m(A(\vec{x})),s,s')$.
Thus $M_h,v[s/s_h] \models
Poss(A(\vec{x}),s)$ if and only if there exists $s_l'$ such that 
$M_l,v[s/s_l,s'/s_l'] \models Do(m(A(\vec{x})),s,s')$.
By the way $B$ is defined, $\langle do([\vec{\alpha},A(\vec{x})],S_0)^{M_h,v}, s_l' \rangle \in B$ if and only if $M_l,v[s/s_l,s'/s_l'] \models
Do(m(A(\vec{x})),s_l,s_l')$ (note that we have standard names and
domain closure for objects and actions, so we can always ground $\vec{x}$).
Thus conditions (2) and (3) hold for $\langle s_h, s_l \rangle =
\langle S_0^{M_h}, S_0^{M_l}\rangle$.
\\
Induction step: Assume that if $\langle s_h, s_l \rangle \in B$ and
the number of actions in $s_h$ is no greater than $n$, then
$\langle s_h, s_l \rangle$ satisfies the three conditions in the
definition of $m$-bisimulation.  We have to show that this must also
hold for any $\langle s_h, s_l \rangle \in B$ where $s_h$ contains
$n+1$ actions.
First we show that condition 1  in the definition of $m$-bisimulation holds.
If $\langle s_h, s_l \rangle \in B$ and $s_h$ contains $n+1$ actions,
then due to the way $B$ is defined, there exist situations $s_h'$ and $s_l'$,
a ground high-level action sequence $\vec{\alpha}$ of length $n$, and a
ground high-level action $A(\vec{c})$, such that 
$s_h = do(A(\vec{c}), do(\vec{\alpha},S_0))^{M_h}$, 
$s_h' = do(\vec{\alpha},S_0)^{M_h}$, 
$M_l,v[s/s_l'] \models Do(m(\vec{\alpha}),S_0,s)$, and 
$\langle s_h', s_l' \rangle \in B$.
$s_h'$ contains $n$ actions so 
by the induction hypothesis, $\langle s_h', s_l' \rangle$ satisfies
the three conditions in the definition of $m$-bisimulation,
in particular $s_h' \simeq_m^{M_h,M_l}  s_l'$. 
By Lemma \ref{lem:sitSupBisim}, it follows that $M_h,v[s/s_h'] \models
\phi^{ssa}_{F,A}(\vec{y},\vec{c})[s]$ if and only if $M_l,v[s/s_l'] \models
m(\phi^{ssa}_{F,A}(\vec{y},\vec{c}))[s]$ for any high-level fluent
$\mathit{F} \in \mathit{\F_h}$.  Thus by the successor state axiom for
$F$, $M_h,v[s/s_h'] \models F(\vec{y},do(A(\vec{c}),s))$ if and only
if  $M_l,v[s/s_l'] \models m(\phi^{ssa}_{F,A}(\vec{y},\vec{c}))[s]$.  
By condition (c), we have that 
$M_l,v[s/s_l'] \models m(\phi^{ssa}_{F,A}(\vec{y},\vec{c}))[s]$ if and only if
$M_l,v[s/s_l] \models m(F(\vec{y}))[s]$.  Thus $M_h,v[s/s_h'] \models
F(\vec{y},do(A(\vec{c}),s))$ if and only if $M_l,v[s/s_l]
\models m(F(\vec{y}))[s]$.
Therefore, $s_h \simeq_m^{M_h,M_l}  s_l$, i.e., condition 1  in the 
definition of $m$-bisimulation holds.
\\
We can show that $\langle s_h, s_l \rangle$, where $s_h$
contains $n+1$ actions, satisfies conditions 2 and 3 in the definition of
$m$-bisimulation, by exactly the same argument as in the base case.
\qed

With these lemmas in hand, we can prove our main result:

\noindent 
\textbf{Theorem}~ \ref{thm:verifySound} 
 $\D^h$ is a sound abstraction of $\D^l$ relative to mapping $m$ if and only if
\begin{small}
\begin{description}
\item[(a)]
$\D^l_{S_0} \cup \D^l_{ca} \cup \D^l_{coa} \models m(\phi)$, for all
$\phi \in D^h_{S_0}$,
\item[(b)]
$\mathit{\D^l \cup \C \models \forall s. Do(\anyseqhl,S_0o,s) \limp}$ \\
\hspace*{3em} $\bigwedge_{A_i \in \A^h}
    \forall \vec{x}.
 (m(\phi^{Poss}_{A_i}(\vec{x}))[s] 
    \equiv \exists s' Do(m(A_i(\vec{x})),s,s')),$
\item[(c)]
$\D^l \cup \C \models \forall s. Do(\anyseqhl,S_0,s) \limp $ \\
\hspace*{3em} $\bigwedge_{A_i \in \A^h}  \forall \vec{x}, s'. (Do(m(A_i(\vec{x})),s,s') \limp$ \\
\hspace*{5em} $\bigwedge_{F_i \in \F^h}  \forall \vec{y}  (m(\phi^{ssa}_{F_i,A_i}(\vec{y},\vec{x}))[s] \equiv m(F_i(\vec{y}))[s'])),$
\end{description}
\end{small}%
\noindent
where $\phi^{Poss}_{A_i}(\vec{x})$ is the right hand side (RHS) of the precondition axiom for action $A_i(\vec{x})$, and $\phi^{ssa}_{F_i,A_i}(\vec{y},\vec{x})$ is the RHS of the successor state axiom for $F_i$ instantiated with action
$A_i(\vec{x})$ where action terms have been eliminated using
$\D^h_{ca}$. 

\textbf{Proof}
$(\Rightarrow)$ 
By contradiction. 
Assume that $\D_h$ is a sound abstraction of $\D_l$ wrt $m$.
Suppose that condition (a) does not hold, i.e., there exists
$\phi \in D^h_{S_0}$ such that $\D^l_{S_0} \cup \D^l_{ca} \cup
\D^l_{coa} \not\models m(\phi)$.
Thus there exists a model $M_l'$ of $\D^l_{S_0} \cup \D^l_{ca} \cup
\D^l_{coa}$ such that $M_l' \not\models m(\phi)$.  By the relative
satisfiability theorem for basic action theories \cite{PirriR:JACM99,Reiter01-Book}, this model can be
extended to a model $M_l$ of $\D_l \cup \C$ such that $M_l \not\models m(\phi)$.
Since $\D_h$ is a sound abstraction of $\D_l$ wrt $m$, 
there exists a model $M_h$ of $\D_h$ such that $M_h \sim_m M_l$.
By Theorem \ref{thm:bisimL2HOff}, it follows that $M_h \not\models \phi$.
Thus $\D_h \not\models  D^h_{S_0}$, contradiction.
\\[1ex]
Now suppose that condition (b) does not hold.
Then there exists a model $M_l$ of $\D_l \cup \C$ such that $M_l$
falsifies condition (b).
Since $\D_h$ is a sound abstraction of $\D_l$ wrt $m$, 
there exists a model $M_h$ of $\D_h$ such that $M_h \sim_m M_l$.
But then by Lemma \ref{lem:verifySound1}, $M_l$ must satisfy condition (b),
contradiction.
\\[1ex]
We can prove that condition (c) must hold using Lemma
\ref{lem:verifySound1} by the same argument as for condition (b).
\\[1ex]
$(\Leftarrow)$ Assume that conditions (a), (b), and (c) hold. 
Take a model $M_l$ of $\D^l \cup \C$.  Let $M_h$ be a
model of the high-level language such that
\begin{description}
\item[(i)] $M_h$ has the same object domain as $M_l$ and interprets all
  object terms like $M_l$,
\item[(ii)] $M_h \models \D^h_{ca}$,
\item[(iii)] $M_h \models \Sigma$,
\item[(iv)]  $M_h,v \models F(\vec{x},do(\vec{\alpha},S_0))$ if and only if $M_l,v \models
  \exists s. Do(m(\vec{\alpha}),S_0,s) \land  m(F(\vec{x}))[s]$ for
  all fluents $F \in \F^h$ and all ground high-level action sequences
  $\vec{\alpha}$.
\item [(v)] $M_h \models Poss(A(\vec{x}),do(\vec{\alpha},S_0))$ if and only if
$M_l \models \exists s. Do(m(\vec{\alpha}),S_0,s) \land \exists s' Do(m(A(\vec{x})),s,s')$ 
\end{description}
It follows immediately that $M_h \models \Sigma \cup \D^h_{ca} \cup
\D^h_{coa}$.
By condition (iv) above, we have that $S_0^{M_h} \simeq_m^{M_h,M_l} S_0^{M_l}$. 
Thus by condition (a) and Lemma \ref{lem:sitSupBisim}, 
we have that $M_h \models \D^h_{S_0}$ .
By condition (b) of the Theorem and conditions (iv) and (v) above, $M_h \models
\D^h_{Poss}$.
By condition (c) of the Theorem and condition (iv) above, $M_h \models
\D^h_{ssa}$.
Thus  $M_h \models \D^h$.
\\
Now $M_h$ and $M_l$ satisfy all the conditions for applying Lemma
\ref{lem:verifyComplete1}, by which it follows that  $M_h \sim_m M_l$.
\qed

\noindent
\textbf{Proposition}~ \ref{prop:DhEgSoundAbsElEg}
$\D^{eg}_h$ is a sound abstraction of $\D^{eg}_l$ wrt $m^{eg}$.

\textbf{Proof:} 
We prove this using Theorem \ref{thm:verifySound}. 
\\
(a) It is easy to see that $\D^l_{S_0} \cup \D^l_{ca} \cup \D^l_{coa} \models m(\phi)$, for all
$\phi \in D^h_{S_0}$.
\\
(b)
For the $deliver$ high-level action, we need to show that:
\[\begin{array}{l}
\D^l \cup \C \models Do(\anyseqhl,S_0,s) \limp\\
\quad 
    \forall \mathit{sID}.
 (m(\exists l.Dest_{HL}(\mathit{sID},l,s) \land At_{HL}(\mathit{sID}, l, s))\\
\hspace{4em}    \equiv \exists s' Do(m(deliver(sID)),s,s')),
\end{array}\]
i.e.,
\[\begin{array}{l}
\D^l \cup \C \models Do(\anyseqhl,S_0,s) \limp
    \forall \mathit{sID}.\\
\quad (\exists l.Dest_{LL}(\mathit{sID},l,s) \land At_{LL}(\mathit{sID}, l, s)\\
\quad \,\, \,    \equiv \exists s' Do([unload(\mathit{sID}); getSignature(\mathit{sID})],s,s')).
\end{array}\]
It is easy to check that the latter holds as $\exists l.Dest_{LL}(\mathit{sID},l,s) \land
At_{LL}(\mathit{sID}, l, s)$ is the precondition of $unload(\mathit{sID})$ and
$unload(\mathit{sID})$ ensures that the precondition of
$getSignature(\mathit{sID})$ holds.

For the $takeRoute$ action, we need to show that:
\[\begin{array}{l}
\D^l \cup \C \models Do(\anyseqhl,S_0,s) \limp \forall \mathit{sID}, r, o, d.\\
\quad  m(o \neq d \land At_{HL}(\mathit{sID}, o, s) \land CnRoute_{HL}(r, o, d,s)
    \land {}\\
\hspace{3em}    (r=Rt_B \limp \neg Priority(\mathit{sID},s)))\\
\qquad    \equiv \exists s' Do(m(takeRoute(\mathit{sID}, r, o, d)),s,s'),
\end{array}\]
i.e.,
\[\begin{array}{l}
\D^l \cup \C \models Do(\anyseqhl,S_0,s) \limp
    \forall \mathit{sID}, r, o, d.\\
\quad o \neq d \land At_{LL}(\mathit{sID}, o, s) \land CnRoute_{LL}(r, o, d,s) \land {}\\
\hspace{3em}    (r=Rt_B \limp \neg (BadWeather(s) \lor Express(\mathit{sID},s)))\\
\qquad    \equiv \exists s' Do(m(takeRoute(\mathit{sID}, r, o, d)),s,s').
\end{array}\]
It is easy to show that the latter holds as the left hand side of the
$\equiv$ is equivalent to $m(takeRoute(\mathit{sID}, r, o, d))$ being
executable in $s$.
First, we can see that the left hand side of the
$\equiv$ is equivalent to the preconditions of first $takeRoad$ action
in $m(takeRoute(\mathit{sID}, r, o, d))$, noting that in the case where $r=Rt_B$,
$takeRoute(\mathit{sID}, r, o, d)$ is mapped into $takeRoad$ to destination
$L3$, and that the only road that is closed is $Rd_e$; the latter can
be proved to always remain true by induction on situations.
Moreover, the preconditions of the second $takeRoad$ action
in $m(takeRoute(\mathit{sID}, r, o, d))$ must hold given this and that the first
$takeRoad$ has occurred.
\\
(c) For the high-level action $deliver$ we must show that:
\[\begin{array}{l}
\D^l \cup \C \models Do(\anyseqhl,S_0,s) \limp \\
\quad \forall \mathit{sID},s'. (Do(m(deliver(\mathit{sID})),s,s') \limp\\
\quad \quad \bigwedge_{F_i \in \F^h}  \forall \vec{y}   (m(\phi^{ssa}_{F_i,deliver}(\vec{y},\mathit{sID}))[s] \equiv m(F_i(\vec{y}))[s'])).
\end{array}\]
For the high-level fluent $Delivered$, we must show that
\[\begin{array}{l}
\D^l \cup \C \models Do(\anyseqhl,S_0,s) \limp \\
\quad \forall \mathit{sID},s'. (Do(m(deliver(\mathit{sID})),s,s') \limp\\
\quad \quad \forall \mathit{sID'}   ((\mathit{sID'}=\mathit{sID} \lor Unloaded(\mathit{sID'},s') \land
    Signed(\mathit{sID'},s')) \equiv \\
    \hspace{5em} Unloaded(\mathit{sID'},s') \land
    Signed(\mathit{sID'},s')).
\end{array}\]
This is easily shown given that $m^{eg}(deliver(\mathit{sID})) = unload(\mathit{sID});$ \linebreak
$getSignature(\mathit{sID})$, using the successor state axioms for $Unloaded$
and $Signed$.
For the other high-level fluents, the result follows easily as \linebreak
$m^{eg}(deliver(\mathit{sID}))$ does not affect their refinements.
\\
For the action $takeRoute$ we must show that:
\[\begin{array}{l}
\D^l \cup \C \models Do(\anyseqhl,S_0,s) \limp \\
\quad \forall \mathit{sID}, r, o, d,s'. (Do(m(takeRoute(\mathit{sID},r,o,d)),s,s') \limp\\
\quad \quad \bigwedge_{F_i \in \F^h}  \forall \vec{y}
    (m(\phi^{ssa}_{F_i,takeRoute}(\vec{y},\mathit{sID}, r, o, d))[s]
    \equiv m(F_i(\vec{y}))[s'])).
\end{array}\]
For the high-level fluent $At_{HL}$, we must show that
\[\begin{array}{l}
\D^l \cup \C \models Do(\anyseqhl,S_0,s) \limp \\
\quad \forall \mathit{sID}, r, o, d,s'. (Do(m(takeRoute(\mathit{sID},r,o,d)),s,s') \limp\\
\qquad \forall \mathit{sID}',l.(At_{LL}(\mathit{sID'},l,s')) \equiv \\
\qquad \quad (\mathit{sID}' = \mathit{sID} \land l = d) \lor {}\\
\qquad \quad At_{LL}(\mathit{sID'}, l, s) \land 
\neg (\mathit{sID}' = \mathit{sID} \land o = l)).
\end{array}\]
This is easily shown given how $takeRoute$ is refined by $m^{eg}$,
 using the successor state axioms for $At_{LL}$.
For the other high-level fluents, the result follows easily as
$m^{eg}(takeRoute(\mathit{sID},r,o,d))$ does not affect their refinements.
\qed

\subsection{Complete Abstraction}



\noindent
\textbf{Theorem} \ref{Thm:ExecutableL2HOffEntailComp}
Suppose that $\D_h$ is a complete abstraction of $\D_l$ relative to
mapping $m$.  Then for any ground high level action sequence $\vec{\alpha}$ and
any high level situation suppressed formula $\phi$, if
$\D_l \cup \C \models \exists s. Do(m(\vec{\alpha}),S_0,s) \land
m(\phi)[s]$,
then
$\D_h \models Executable(do(\vec{\alpha},S_0)) \land \phi[do(\vec{\alpha},S_0)]$.

\textbf{Proof:}
Assume that $\D_h$ is a complete abstraction of $\D_l$ wrt $m$ and that  
$\D_l \cup \C \models \exists s. Do(m(\vec{\alpha}),S_0,s) \land
m(\phi)[s]$.
Take an arbitrary model $M_h$ of $\D_h$.  Since $\D_h$ is a complete
abstraction of $\D_l$ wrt $m$, there exists a model $M_l$ of $\D_l
\cup \C$ such that $M_h \sim_m M_l$.
It must be the case that $\M_l \models \exists s. Do(m(\vec{\alpha}),S_0,s) \land
m(\phi)[s]$.
Therefore by Theorem \ref{thm:bisimL2HOff}, we must also have that
$\M_h \models Executable(do(\vec{\alpha},S_0)) \land
\phi[do(\vec{\alpha},S_0)]$.
Since  $M_h$ was an arbitrarily chosen model of $\D_h$, the thesis follows.
\qed

\noindent
\textbf{Theorem} \ref{thm:verifySoundComplete}~~
If $\D^h$ is a sound abstraction of $\D^l$ wrt mapping $m$, then
$\D^h$ is also a complete abstraction of $\D^l$ wrt mapping $m$
if and only if
for every model $M_h$ of $\D^h_{S_0} \cup \D^h_{ca} \cup \D^h_{coa}$,
there exists a model  $M_l$ of $\D^l_{S_0} \cup \D^l_{ca}
\cup\D^l_{coa}$ such that $S_0^{M_h} \simeq_m^{M_h,M_l} S_0^{M_l}$.

\textbf{Proof:}
Assume that $\D^h$ is a sound abstraction of $\D^l$ wrt mapping $m$.\\
($\Rightarrow$)
Suppose that $\D^h$ is a complete abstraction of $\D^l$ wrt mapping
$m$.
Take an arbitrary model of $M_h$ of $\D^h_{S_0} \cup \D^h_{ca} \cup
\D^h_{coa}$.
By the relative satisfiability theorem for basic action theories of
\cite{PirriR:JACM99,Reiter01-Book}, $M_h$ can be extended to satisfy all of $\D^h$.
Since $\D^h$ is a complete abstraction of $\D^l$ wrt $m$, by 
definition, there exists a model $M_l$ of $\D_l$ such that  $M_l
\sim_m M_h$.
It follows by the definition of $m$-bisimulation that $S_0^{M_h} \simeq_m^{M_h,M_l} S_0^{M_l}$.\\
($\Leftarrow$)
Suppose that for every model $M_h$ of $\D^h_{S_0} \cup \D^h_{ca} \cup \D^h_{coa}$,
there exists a model  $M_l$ of $\D^l_{S_0} \cup \D^l_{ca}
\cup\D^l_{coa}$ such that $S_0^{M_h} \simeq_m^{M_h,M_l} S_0^{M_l}$.
Take an arbitrary model $M_h$ of $\D^h$.
Since $M_h$ is also a model of $\D^h_{S_0} \cup \D^h_{ca} \cup \D^h_{coa}$,
then there exists a model  $M_l$ of $\D^l_{S_0} \cup \D^l_{ca}
\cup\D^l_{coa}$ such that $S_0^{M_h} \simeq_m^{M_h,M_l} S_0^{M_l}$.
Clearly, $M_l$ can be extended to satisfy all of $\D^l$ by the relative satisfiability theorem for basic action theories of \cite{PirriR:JACM99,Reiter01-Book}.
Moreover, $M_l$ can be extended to satisfy $\C$ (by the results in \cite{DBLP:journals/ai/GiacomoLL00}).
Since $\D^h$ is also a sound abstraction of $\D^l$ wrt $m$, by
Theorem \ref{thm:verifySound} it follows that:
\[\begin{array}{l}
M_l \models Do(\anyseqhl,S_0,s) \limp \\
\hspace*{4.5em} \bigwedge_{A_i \in \A^h}
    \forall \vec{x}.
 (m(\phi^{Poss}_{A_i}(\vec{x}))[s] 
    \equiv \exists s' Do(m(A_i(\vec{x})),s,s'))\\
\mbox{and}\\
\mathit{M_l \models Do(\anyseqhl,S_0,s) \limp}  \\
\hspace*{4.5em} \bigwedge_{A_i \in \A^h}  \forall \vec{x}, s'. (Do(m(A_i(\vec{x})),s,s') \limp \\
\hspace*{7.5em} \bigwedge_{F_i \in \F^h}  \forall \vec{y}
(m(\phi^{ssa}_{F_i,A_i}(\vec{y},\vec{x}))[s] \equiv
m(F_i(\vec{y}))[s'])),
\end{array}\]
where $\phi^{Poss}_{A_i}(\vec{x})$ and
$\phi^{ssa}_{F_i,A_i}(\vec{y},\vec{x})$ are as in Theorem \ref{thm:verifySound}.
Thus by Lemma \ref{lem:verifyComplete1}, it follows that $M_h \sim_m M_l$.
Thus $\D^h$ is a complete abstraction of $\D^l$ wrt $m$, by the 
definition of complete abstraction.
\qed

\noindent
\textbf{Theorem} \ref{thm:verifyComplete}
$\D^h$ is a complete abstraction of $\D^l$ relative to mapping $m$ if
and only if
  for every model $M_h$ of $\D^h$,
  there exists a model  $M_l$ of $\D^l \cup \C$ such that
\begin{small}
\begin{description}
\item[(a)]
  $S_0^{M_h} \simeq_m^{M_h,M_l} S_0^{M_l}$, 
\item[(b)]
$\mathit{M_l \models \forall s. Do(\anyseqhl,S_0,s) \limp}$ \\
\hspace*{0.5em} $\bigwedge_{A_i \in \A^h}
    \forall \vec{x}.
 (m(\phi^{Poss}_{A_i}(\vec{x}))[s] 
    \equiv \exists s' Do(m(A_i(\vec{x})),s,s'))$,
\item[(c)] 
$M_l \models \forall s. Do(\anyseqhl,S_0,s) \limp $ \\
\hspace*{2em} $\bigwedge_{A_i \in \A^h}  \forall \vec{x}, s'. (Do(m(A_i(\vec{x})),s,s') \limp$ \\
\hspace*{3.3em} $\bigwedge_{F_i \in \F^h}  \forall \vec{y}
(m(\phi^{ssa}_{F_i,A_i}(\vec{y},\vec{x}))[s] \equiv
m(F_i(\vec{y}))[s']))$,


\end{description}
\end{small}%

\noindent
where $\phi^{Poss}_{A_i}(\vec{x})$ and $\phi^{ssa}_{F_i,A_i}(\vec{y},\vec{x})$ are as in Theorem \ref{thm:verifySound}.


\textbf{Proof:}
($\Rightarrow$)
Suppose that $\D^h$ is a complete abstraction of $\D^l$ wrt mapping
$m$.
Take an arbitrary model of $M_h$ of $\D^h$.
Since $\D^h$ is a complete abstraction of $\D^l$ wrt $m$, by 
definition, there exists a model $M_l$ of $\D_l$ such that  $M_l
\sim_m M_h$.
It follows by the definition of $m$-bisimulation that $S_0^{M_h}
\simeq_m^{M_h,M_l} S_0^{M_l}$.
Furthermore, by Lemma \ref{lem:verifySound1}, it follows that:
\[\begin{array}{l}
M_l \models \forall s. Do(\anyseqhl,S_0,s) \limp \\
\hspace*{4.5em} \bigwedge_{A_i \in \A^h}
    \forall \vec{x}.
 (m(\phi^{Poss}_{A_i}(\vec{x}))[s] 
    \equiv \exists s' Do(m(A_i(\vec{x})),s,s'))\\
\mbox{and}\\
M_l \models \forall s. Do(\anyseqhl,S_0,s) \limp  \\
\hspace*{4.5em} \bigwedge_{A_i \in \A^h}  \forall \vec{x}, s'. (Do(m(A_i(\vec{x})),s,s') \limp \\
\hspace*{7.5em} \bigwedge_{F_i \in \F^h}  \forall \vec{y}
    (m(\phi^{ssa}_{F_i,A_i}(\vec{y},\vec{x}))[s] \equiv
    m(F_i(\vec{y}))[s']))
  \end{array}\]
where $\phi^{Poss}_{A_i}(\vec{x})$ and
$\phi^{ssa}_{F_i,A_i}(\vec{y},\vec{x})$ are as in Theorem \ref{thm:verifySound}.
\\
($\Leftarrow$)
The thesis follows immediately from Lemma \ref{lem:verifyComplete1} and the definition of
complete abstraction.
\qed

 \textbf{Corollary} \ref{cor:complete-sound-then-completeabs}
  If $\D^h_{S_0}$ is a complete theory
  (i.e., for any situation suppressed formula $\phi$,
  either $\D^h_{S_0} \models   \phi[S_0]$ or
   $\D^h_{S_0} \models   \lnot \phi[S_0]$)
   and  $\D^l$ is satisfiable,
   then if $\D^h$ is a sound abstraction of $\D^l$ wrt $m$,
   then $\D^h$ is also a complete but abstraction of $\D^l$ wrt $m$.

\textbf{Proof:}
$\D^l$ is satisfiable, so it has a model $M_l$, and since $\D^h$ is a
sound abstraction of $\D^l$ wrt $m$, $\D^h$ has a model $M_h$ such that
$M_h \sim_m M_l$.  By the definition of $m$-bisimilar model, this
implies that $S_0^{M_h} \simeq_m^{M_h,M_l} S_0^{M_l}$.
Take an arbitrary model $M_h'$ of $\D_h$.  Since
$\D^h_{S_0}$ is a complete theory, we have that
$M_h' \models \phi[S_0]$ iff $M_h \models \phi[S_0]$
for all situation suppressed formulas $\phi$.
It follows that  $S_0^{M_h'} \simeq_m^{M_h,M_l} S_0^{M_l}$.
Then by Theorem \ref{thm:verifySoundComplete}.
we have that $\D^h$ is a complete abstraction of $\D^l$ wrt $m$.
\qed

\subsection{Monitoring and Explanation}

\textbf{Theorem} \ref{thm:unique-lp_m}
  For any refinement mapping $m$ from $\D_h$ to $\D_l$, we have that:
\begin{enumerate}
\item
$D_l \cup \C \models \forall s.\exists s'. lp_m(s,s')$,
\item
$\D_l \cup \C \models \forall s \forall s_1 \forall s_2. lp_m(s,s_1)
\land lp_m(s,s_2) \supset s_1 = s_2$.
\end{enumerate}

\textbf{Proof:} (1) We have that
$\D_l \cup \C \models Do(\mathit{\anyseqhl},S_0,S_0)$ since $\mathit{\anyseqhl}$ is a
nondeterministic iteration that can execute 0 times.  So even if there
is no $s''$ such that $S_0 < s'' \leq s \land Do(\mathit{\anyseqhl},S_0,s'')$, the
result holds.
\\
(2) Take an arbitrary model $M_l$ of $\D_l \cup \C $ and assume that
$M_l, v \models lp_m(s,s_1) \land lp_m(s,s_2)$.
We have that
$\D_l \cup \C \models lp_m(s,s') \supset s' \leq s$.  Moreover, we have a total
ordering on situations $s'$ such that $s' \leq s$.  If $M_l, v \models s_1 < s_2$,
then $s_1$ can't be the largest prefix of $s$ that can be produced by
executing a sequence of high-level actions, and we can't have
$M_l, v \models lp_m(s,s_1)$.  Similarly if $M_l, v \models s_2 < s_1$, we can't have
$M_l, v \models lp_m(s,s_2)$.  It follows that $M_l, v \models s_1 = s_2$.
\qed

\noindent
\textbf{Theorem} \ref{thm:unique-m_inv}
  Suppose that we have a refinement mapping $m$ from $\D_h$ to $\D_l$
  and that Constraint \ref{asm:firstThree}
  holds. Let
  $M_l$ be a model of $\D_l \cup \C$. Then for any ground situation
  terms $S_s$ and $S_e$ such that
  $M_l \models Do(\mathit{\anyseqhl},S_s, S_e)$, there exists a unique
  ground high-level action sequence $\vec{\alpha}$ such that
  $M_l \models Do(m(\vec{\alpha}),S_s,S_e)$.

\textbf{Proof:}
Since, $M_l \models Do(\mathit{\anyseqhl},S_s,S_e)$, there exists a $n\in
\Nat$ such that $M_l \models Do(\mathit{\anyonehl}^n,S_0,S)$.
Since we have standard names for objects, it follows that there exists
a ground high-level action sequence $\vec{\alpha}$ such that $M_l \models
Do(m(\vec{\alpha}),S_s,S_e)$.
Now let's show by induction on the length of $\vec{\alpha}$
that there is no ground high-level action sequence
$\vec{\alpha}' \neq \vec{\alpha}$ such that $M_l \models
Do(m(\vec{\alpha}'),S_s,S_e)$.
Base case $\vec{\alpha} = \epsilon$: Then  $M_l \models
Do(m(\vec{\alpha}),S_s,S_e)$ implies $M_l \models S_s = S_e$ and there
is no $\vec{\alpha}' \neq \epsilon$ such that $M_l \models
Do(m(\vec{\alpha}'),S_s,S_e)$, since by Constraint \ref{asm:firstThree}(c)
$\D_l \cup \C \models Do(m(\beta),s,s') \supset s < s'$ for any ground
high-level action term $\beta$.
Induction step: 
Assume that the claim holds for any $\vec{\alpha}$ of length $k$.
Let's show that it must hold for any $\vec{\alpha}$ of length $k+1$.
Let $\vec{\alpha} = \beta\vec{\gamma}$.  There exists $S_i$ such that
$Do(m(\beta),S_s,S_i) \land S_i \leq S_e$. By Constraint
\ref{asm:firstThree}(a), there is no $\beta' \neq \beta$ and $S_i'$ such that
$Do(m(\beta'),S_s,S_i') \land S_i' \leq S_e$.  
By Constraint \ref{asm:firstThree}(b), there is no  $\S_i' \neq S_i$
such that $Do(m(\beta),S_0,S_i') \land S_i' \leq S$.
Then by the induction hypothesis, there is no ground high-level action sequence
$\vec{\gamma}' \neq \vec{\gamma}$ such that $M_l \models
Do(m(\vec{\gamma}'),S_i,S_e)$.
\qed

\noindent

\textbf{Theorem} \ref{thm:lp_m_rest}
If  $m$ is a refinement mapping from $\D_h$ to $\D_l$ and 
Constraint \ref{asm:offMatchLL} holds, then we have that:
\[\begin{array}{l}
\D_l \cup \C \models \forall s, s'. Executable(s) \land lp_m(s,s')
\supset \\
\hspace{6.6em}\exists \delta.Trans^*(\mathit{\anyonehl},s',\delta,s)
\end{array}\]

\textbf{Proof:}
Take an arbitrary model $M_l$ of $\D_l \cup \C $ and assume that \linebreak
$M_l, v \models Executable(s) \land lp_m(s,s')$.
Since $M_l, v \models lp_m(s,s')$, we have that
$M_l, v \models Do(\mathit{\anyseqhl},S_0,s')$ and thus that $M_l, v \models Trans^*(\mathit{\anyseqhl}, $ \linebreak $S_0,\mathit{\anyseqhl},s')$.  
Since $M_l, v \models Executable(s)$, by Constraint
\ref{asm:offMatchLL} we have that
$M_l, v \models \exists \delta.Trans^*(\mathit{\anyseqhl},S_0,\delta,s)$.
Thus, it follows that $M_l, v \models \exists \delta.Trans^*(\mathit{\anyonehl},s',\delta,s)$.
\qed

\section{Additional Details on \ConGolog Semantics}
The definitions of $Trans$ and $Final$ for the \ConGolog constructs used in this paper are as in
\cite{DBLP:conf/kr/GiacomoLP10}. Note that since $Trans$ and  $Final$ take programs (that include tests of formulas) as arguments, this requires encoding formulas and programs as terms; see  \cite{DBLP:journals/ai/GiacomoLL00} for details.

The predicate $Trans$ is characterized by the following
set of axioms:  

\begin{small}
\[\begin{array}{l}
    Trans(\alpha,s,\delta',s') \equiv
s'=do(\alpha,s) \land \Poss(\alpha,s) \land \delta'=True? \\[0.5ex]
Trans(\varphi?,s,\delta',s') \equiv False \\[0.5ex]
Trans(\delta_1; \delta_2,s,\delta',s') \equiv{} \\[0.5ex]
\qquad
	Trans(\delta_1,s,\delta_1',s') \land \delta'=\delta_1';\delta_2 \lor{}\\
\qquad
		Final(\delta_1,s) \land Trans(\delta_2,s,\delta',s') \\[0.5ex]
    Trans(\delta_1 \ndet \delta_2,s,\delta',s') \equiv
	Trans(\delta_1,s,\delta',s') \lor Trans(\delta_2,s,\delta',s') \\[0.5ex]
Trans(\pi x.\delta, s, \delta',s') \equiv 
	\exists x. Trans(\delta,s,\delta',s')\\[0.5ex]
Trans(\delta^*, s, \delta',s') \equiv
	Trans(\delta,s,\delta'',s')\land \delta'=\delta'';\delta^*\\
Trans(\delta_1 \conc \delta_2,s,\delta',s') \equiv{} \\ 
\qquad
	Trans(\delta_1,s,\delta_1',s')  \land \delta'=\delta_1'\conc\delta_2 \lor {}\\
\qquad
		Trans(\delta_2,s,\delta_2',s')  \land \delta'=\delta_1\conc\delta_2' 
\end{array}\]
\end{small}%

The predicate $Final$ is characterized by the following
set of axioms: 

\begin{small}
\[\begin{array}{l}
Final(\alpha,s) \equiv False \\
Final(\varphi?,s) \equiv \varphi[s] \\
Final(\delta_1; \delta_2,s) \equiv 
        Final(\delta_1,s) \land Final(\delta_2,s)\\
Final(\delta_1 | \delta_2,s) \equiv 
	Final(\delta_1,s) \lor Final(\delta_2,s) \\
Final(\pi x.\delta, s) \equiv
	\exists x. Final(\delta,s)\\
Final(\delta^*, s) \equiv True\\
Final(\delta_1 \conc \delta_2,s) \equiv 
       Final(\delta_1,s) \land Final(\delta_2,s)
\end{array}\]
\end{small}%
These are in fact the usual ones
\cite{DBLP:journals/ai/GiacomoLL00}, except that, following
\cite{DBLP:conf/kr/ClassenL08}, the test construct $\varphi?$ does not yield
any transition, but is final when satisfied.  Thus, it is
a \emph{synchronous} version of the original test construct (it does
not allow interleaving).

Also, note that the construct \textbf{if} $\phi$ \textbf{then} $\delta_1$
\textbf{else} $\delta_2$ \textbf{endIf} is defined as $[\phi?;
\delta_1 ] \mid [\neg \phi?; \delta_2]$ and 
\textbf{while} $\phi$ \textbf{do} $\delta$ \textbf{endWhile}
is defined as $(\phi: \delta)^*; \neg \phi?$.






\end{document}